\documentclass[12pt,preprint2,natbib,iop]{emulateapj} 
\usepackage{graphicx,epsfig,amsmath,multirow,mathrsfs} 
\usepackage{mathtools}
\usepackage{color}

\lefthead{Maseda et al.}
\righthead{Number Density of EELGs in 3D-HST}
\slugcomment{Version: \today}
\begin{document}

\newcommand{\kms}{\>{\rm km}\,{\rm s}^{-1}}
\newcommand{\reff}{r_{\rm{eff}}}
\newcommand{\msol}{M_{\odot}}
\newcommand{\zsol}{Z_{\odot}}
\newcommand{\inverse}[1]{{#1}^{-1}}
\newcommand{\invvar}{\inverse{C}}
\newcommand{\dd}{{\rm d}}
\newcommand{\cgs}{erg~s^{-1}~cm^{-2}}

\title{The Number Density Evolution of Extreme Emission Line Galaxies in 3D-HST: Results from a Novel Automated Line Search Technique for Slitless Spectroscopy\altaffilmark{*}}
\author{Michael V. Maseda\altaffilmark{1,2}, Arjen van der Wel\altaffilmark{2}, Hans-Walter Rix\altaffilmark{2}, Ivelina Momcheva\altaffilmark{3}, Gabriel B. Brammer\altaffilmark{3}, Marijn Franx\altaffilmark{1}, Britt F. Lundgren\altaffilmark{4}, Rosalind E. Skelton\altaffilmark{5}, Katherine E. Whitaker\altaffilmark{6}}

\affil{$^1$ Leiden Observatory, P.O. Box 9513, 2300 RA, Leiden, The Netherlands; maseda@strw.leidenuniv.nl}
\affil{$^2$ Max-Planck-Institut f\"ur Astronomie, K\"onigstuhl 17, 69117
Heidelberg, Germany}
\affil{$^3$ Space Telescope Science Institute, 3700 San Martin Dr., Baltimore, MD 21218, USA}
\affil{$^4$ Department of Physics, University of North Carolina Asheville, One University Heights, Asheville, NC 28804, USA}
\affil{$^5$ South African Astronomical Observatory, P.O. Box 9, Observatory, 7935, South Africa}
\affil{$^6$ Department of Physics, University of Connecticut, 2152 Hillside Road, Unit 3046, Storrs, CT 06269, USA}

\altaffiltext{*}{This work is based on observations taken by the 3D-HST Treasury Program and the CANDELS Multi-Cycle Treasury Program with the NASA/ESA HST, which is operated by the Association of Universities for Research in Astronomy, Inc., under NASA contract NAS5-26555.}
\begin{abstract}
The multiplexing capability of slitless spectroscopy is a powerful asset in creating large spectroscopic datasets, but issues such as spectral confusion make the interpretation of the data  challenging. Here we present a new method to search for emission lines in the slitless spectroscopic data from the 3D-HST survey utilizing the Wide-Field Camera 3 on board the \textit{Hubble Space Telescope}.  Using a novel statistical technique, we can detect compact (extended) emission lines at 90\% completeness down to fluxes of 1.5 (3.0)$\times 10^{-17}~\cgs$, close to the noise level of the grism exposures, for objects detected in the deep ancillary photometric data.  Unlike previous methods, the Bayesian nature allows for probabilistic line identifications, namely redshift estimates, based on secondary emission line detections and/or photometric redshift priors.  As a first application, we measure the comoving number density of Extreme Emission Line Galaxies (restframe [O III]$\lambda$5007 equivalent widths in excess of 500~\AA).   We find that these galaxies are nearly 10$\times$ more common above $z\sim1.5$ than at $z\lesssim0.5$.  With upcoming large grism surveys such as \textit{Euclid} and \textit{WFIRST} as well as grisms featuring prominently on the NIRISS and NIRCam instruments on \textit{James Webb Space Telescope}, methods like the one presented here will be crucial for constructing emission line redshift catalogs in an automated and well-understood manner.
\end{abstract}

\section{INTRODUCTION} 


In recent years, combinations of deep imaging and spectroscopy with the \textit{Hubble Space Telescope} (HST) have been used to tackle a number of  outstanding questions in observational astronomy.  The HST has a particular advantage in the near-IR with the Wide-Field Camera 3 (WFC3), due to the lower sky background levels compared to ground-based observatories and the higher spatial resolution.  Using the grism, all sources within the $\sim$ 2$'\times$2$'$ WFC3/IR field of view have dispersed spectra, which are essentially a series of monochromatic (overlapping) two-dimensional images shifted on the detector according to their wavelength.  The combined spatial and spectral information gives insight into e.g. the growth of disks and bulges at high-redshifts \citep{patel,vd13, hathi}, the spatial distribution of star formation \citep{nelson}, the regulation star-formation in massive galaxies \citep{ferreras}, and the role of environment and mergers in shaping the galaxy population \citep{schmidt}.  Additionally, these surveys are very efficient at covering large areas with a superior multiplexing capacity compared to even the most advanced multi-object spectrographs, allowing for complete studies of rare objects such as cold brown dwarfs \citep{masters12} or $4 < z < 7$ Lyman-break and Lyman-$\alpha$-emitting galaxies \citep{pirzkal,rhoads,malhotra}.

That being said, slitless grism spectroscopic data are more complex to interpret than standard spectroscopic data.  Contamination from unrelated spectra makes a detailed analysis of individual objects challenging, particularly in crowded fields, and often only sources detected via shallow ancillary imaging are analyzed, limiting the potential for discovery.  As emission lines contain so much astrophysically-interesting information and are the easiest spectral features to detect in faint sources, their detection tends to be the primary focus of grism surveys.  Different methods for their discovery have been developed and tuned to the various strengths of the specific set of observations.  

\citet{meurer} outline two techniques for finding emission lines in a semi-automated fashion.  The first method relies on the detection of sources in the direct image.  Each source has its corresponding grism spectrum extracted and emission lines are detected by visual inspection.  This is the preferred method of the \textit{WISP} survey \citep{wisp}, a pure-parallel survey using WFC3.  There, spectral extractions are performed using the \textit{aXe} software \citep{axe} developed to analyze HST grism data.  Most spectra are taken with both the $G102$ and $G141$ grisms, covering an effective wavelength range of $0.8 - 1.7~ \mu m$.

The second method involves searching for emission lines directly in the grism frames.  This is done by smoothing the grism frame with a sausage-shaped filter, designed to match the spatial extent of dispersed first-order spectrum, and then subtracting this smoothed image from the original frame.  This effectively removes continuum sources from the image while leaving spectrally-compact features, such as emission lines, which can be detected using a simple signal-to-noise (S/N) cut.  Undispersed zeroth order spectra from the brightest objects also appear as point sources, but their position is well-known and they are easily masked.  For each detected feature in this subtracted frame, a cutout of the direct image is inspected to determine which source could have produced the feature.  This is the preferred method of the \textit{PEARS} survey \citep{straughn,pears} using the $G800L$ grism of the ACS covering $0.5 - 1.1~\mu m$.  That survey has the added advantage of having multiple position angles (PAs), such that identification of the source of the emission line in the direct image is simply identifying the area where the different spectral traces for the same feature intersect.

Both approaches have a serious limitation: while the line candidate identification may be semi-automated the significance assessment is not, but instead relies on visual inspection, often requiring multiple people to grade the reliability of each line.  Such an approach makes a determination of the true completeness difficult and is not feasible for larger data sets.  In addition, the first approach only has high fidelity in redshift assignments in the cases where more than one line is detected.  We jointly analyze photometric redshift information and grism spectra in the determination of redshifts. This differs from the approach by some of the aforementioned studies that utilize independent photometric redshifts to verify their results.  Relying only on multiple line detections introduces problems in the subjective nature of line identification as well as preferentially selecting objects in certain redshift ranges, typically where both H$\alpha$ and [O III]$\lambda\lambda4959,5007$ are visible.  Indeed, while the quoted flux limit for compact emission lines in \textit{WISP} is 5 $\times$ 10$^{-17} \cgs$, which is based on the WFC3 exposure time calculator, these lines are often only detected in objects that have a second, brighter line.

In this paper, we develop and apply a new method for detecting significant emission features in grism spectroscopic data, using data from the 3D-HST survey\footnote{http://3dhst.research.yale.edu/} \citep{vd,gb}.  3D-HST is a near-infrared spectroscopic Treasury program utilizing WFC3. This program provides WFC3/IR primary and Advanced Camera for Surveys (ACS) parallel imaging and grism spectroscopy over approximately three-quarters ($\sim$700 square arcminutes) of the CANDELS fields \citep{candels1,candels2}.  We focus here on the WFC3 grism data, which utilizes the $G141$ grism covering 1075 to 1700 nm.

3D-HST provides several advantages over other existing HST grism surveys.  As the observations are dithered in a four-point pattern, the processed images offer additional robustness against the effect of hot and bad pixels that a pure-parallel survey cannot.   We also have the ability to interlace the frames instead of drizzling them, where the pixels from the input images are alternately placed in the output image according to the position of the pixel centers in the original images: see Figure 3 of \citet{gb}.  Interlacing the frames results in better noise characteristics, which is crucial to consider when pushing towards the faint limits of emission line sensitivity; the interlacing procedure is described fully in \citet{Momcheva}.  This also provides higher spatial resolution than drizzling (each interlaced pixel is 23 \AA $~\times$ 0.06$''$) and the ability to more easily identify point-like emission sources.


The remainder of this paper is organized as follows.  In Section \ref{sec:data} we describe the 3D-HST data used.  Section \ref{sec:method} lays out our new method to search for significant emission lines in slitless spectroscopic data, utilizing photometric information as a redshift/emission line position prior.  Extensive tests of the method to determine the completeness function as well as contamination are performed and discussed in Section \ref{sec:completeness}.  We apply our method to the 3D-HST data set in Section \ref{sec:results} to obtain a sample of high-equivalent width (high-EW) emission line galaxies.  We then employ a Bayesian analysis to constrain a parameterized model for their luminosity and redshift distribution.  Finally, in Section \ref{sec:summary}, we summarize and compare these results to previous studies of high-EW galaxies, namely the population of extreme emission line galaxies (EELGs).  We adopt a $\Lambda$CDM cosmology with $\Omega_m=0.3$ and H$_0=70 ~$km s$^{-1}$Mpc$^{-1}$ throughout.

\section{DATA}
\label{sec:data}
The spectroscopic data comes from the 3D-HST survey, designed to provide spectroscopy for four well-studied CANDELS extragalactic fields: \textit{AEGIS, COSMOS, GOODS-S}, and \textit{UDS}.  The data set is supplemented by grism spectroscopy for \textit{GOODS-N} (PI: B. Weiner). (Undispersed) objects are detected in a combined CANDELS/3D-HST $F125W$+$F140W$+$F160W$ image and multiband photometry is obtained as part of the \citet{skelton} photometric catalog.  A thorough description of the 3D-HST spectroscopic release, which includes extracted 2D grism spectra for all $\sim250,000$ objects in the \citet{skelton} photometric catalog, and data reduction methods are given in \citet{Momcheva}.  We briefly summarize some of the important points here.

A model for the grism spectra of the entire field is created as follows.  For a given object, we distribute the light (and consequently the extraction weight in the spatial direction) according to the \textit{EAZY} \citep{eazy} continuum fit at the photometric redshift estimate, with the spatial extent according to the $F125W$+$F140W$+$F160W$ ``postage stamp" image of the object.  Next, for bright objects ($m_{F140W} < 22$) we use the the extracted spectrum as a second step to give the continuum the correct shape and to take the brightest emission lines into account. 

Creating a continuum model individually for all objects allows us to construct a model of the flux distribution for the entire field.  This is useful because of spectral confusion due to overlapping unassociated spectra in the grism data.  Since we create the full modeled spectra for each pointing, each extracted 2D spectrum has the modeled spectra from surrounding objects (``contamination") subtracted.  We also subtract an object's own continuum (``model") in order to search for positive residuals, namely emission features.  For the brightest objects, the model does not always subtract cleanly and can lead to spurious detections in neighboring objects, so we also mask any ``contamination'' regions where the model flux exceeds 0.004 e$^{-}$ s$^{-1}$ px$^{-1}$: see Section \ref{sec:contam}.  Likewise, as we are primarily interested in emission lines with high equivalent widths in this study (see Section \ref{sec:results}), we focus on objects with $F140W_{AB}$ magnitudes fainter than 24.


\section{SIMPLE MODEL FITTING OF EMISSION LINES IN 3D-HST}
\label{sec:method}
In this section we set out to devise a straightforward, probabilistic method to detect emission lines and assess their significance algorithmically.
It is based on the following elements: to start, we calculate what the likelihood of the data is, if there was an emission line with amplitude $A$ at a spectral position $\Delta x$ with respect to the undispersed image. To assess the redshift and significance of any detection, $A>0$, we
incorporate three pieces of independent information as prior beliefs. First, we know that in most instances there is either one or no significant emission line in the entire spectrum of an object; conversely, that means that the vast majority of pixels must contain no emission line. Second, 
any emission line flux we detect must not violate the total flux constraints from the corresponding broad-band image. Finally, in many 
instances we have photometric information about probable redshift of the object (``photometric redshift priors"), which also inform our assessment of any emission line detection significance and line identification. Taken together, this results in a joint statement about whether any object has a significant emission line, and if so at which redshift. In other words, line detection and identification are an integrated process. 

\subsection{Line Detection Formalism}
\label{sec:formalism}

We presume the following algorithm to operate on continuum-subtracted spectra, which we obtain as follows:
every object in the \citet{skelton} photometric catalog has a grism spectrum ($S'$) with an associated noise spectrum ($\sigma_{S}$) and a direct $F125W$+$F140W$+$F160W$-combined postage stamp ($I$, of dimensions $x_{max}$ and $y_{max}$).  From each grism spectrum, $S'$, we subtract its continuum flux model and a flux model for all contaminating spectra \cite[see Section \ref{sec:data} and][]{Momcheva} to obtain a spectrum $S$, in which we search for residual emission features.  

As a convolution kernel in the line search, we need to construct an empirical template for the expected spatial distribution of a monochromatic line-image at any given wavelength. We construct that from the undispersed image, $I$, applying a signal-to-noise cut of $2\sigma$ above the background level.  For all sources whose undispersed image has fewer than 20 pixels above this threshold, we instead use the HST $F140W$ Point Spread Function (PSF) scaled to the same flux as the image. This choice is justified, as the area of 20 (interlaced) pixels is approximately the size of a WFC3 PSF.  

Presuming there is a line image of shape $I$, offset by $\Delta x$ along the spectrum and characterized by an amplitude scale factor $A$,
we calculate the likelihood of the data as:
%

\begin{multline}
\tiny \ln \mathscr{L}(\{S\}|A,\Delta x) =\\
 -\frac{1}{2}\sum^{x_{max}}_{x=0} \sum^{y_{max}}_{y=0} \frac{(S_m(x,y|A) - S(x+\Delta x,y))^2}{\sigma_{S}^2(x+\Delta x,y)},
 \label{eqn:likeli}
\end{multline}

where the spectral emission line model is $S_m(x,y|A) = A\times I(x,y)$.  We are dealing with a two-dimensional projection of the three-dimensional spectral information $S(x_{on-sky},y_{on-sky},\lambda_{rest}$). In this context, $\ln \mathscr{L}(\{S\}|A,\Delta x)$ represents a correlation
between $I$ and $S$ in the dispersion direction for different $\Delta x$. Throughout, we use pixel coordinates $\hat{x}$ in the dispersion direction,
which of course reflect different wavelengths, $\hat{\lambda}$.  The maximum value of $\Delta x$ corresponds to the length of $S$, here denoted as $\Delta x_{max}$.  The scale factor $A$ ranges from 0 to 1.  At a given position, $A = 0$ implies that there is no signal present in the spectrum.  Conversely, $A = 1$ corresponds to a position where the entire flux\footnote{The spectral range of the $F125W$+$F140W$+$F160W$ filter combination overlaps with the $G141$ grism such that those filters cover a slightly larger wavelength range than the grism.} of the galaxy is contained in a single emission feature with the same spatial extent as the direct image and is unresolved spectrally (the mean dispersion of the primary spectral order of the G141 grism is 46.5 \AA/pixel or $R = 130$).   We calculate the likelihood at each integer value of $\Delta x$, noting that the FWHM of the WFC3 Point Spread Function (PSF) is 1.1 native pixels at 1.4 $\mu m$.

However, we want the posterior distribution function $p_{posterior}(A|\{S\},\Delta x)$ of the possible line amplitude $A$, not the
likelihood of the data. The two are related via Bayes's Theorem
\begin{equation}
p_{posterior}(A|\{S\},\Delta x) \propto \mathscr{L}(\{S\}|A,\Delta x) \times p_{prior}(A|\Delta x).
\label{eqn:bayes}
\end{equation} 
More specifically, we want the probability that there is a significant line detection, $A>0$ at any given position: 
\begin{equation}
p(A > 0|\{S\},\Delta x) = \int^{1}_{> 0} p_{posterior}(A|\{S\},\Delta x) dA.
\label{eqn:integral}
\end{equation}

Equation \ref{eqn:bayes} requires the prior information on both $A$ and $\Delta x$.  Absent informative photometric redshift information (see Section \ref{sec:photprior}), $p_{prior}(\Delta x)$ is flat: 
$p_{prior}(\Delta x) = x_{max}/\Delta x_{max}$, for the case of one emission line in the entire spectrum.  
This is typically $\sim 1/300$, i.e. $\ll 1$.  It is this prior $p_{prior}(\Delta x)$ that accounts for the fact that we query
 $\Delta x/x_{max}$ independent positions along the spectrum whether there is a significant line flux. When considering only the likelihood
 (Equation \ref{eqn:likeli}), we would expect one spurious ``3$\sigma$-detection'' for every $\sim$200 independent spectral positions, in the absence of any emission line and in the absence of any systematic errors or residuals. The factor  $p_{prior}(\Delta x)$ prevents such false detections,
 as the {\it prior} probability that there is \textit{no} line is $1 - p_{prior}(\Delta x)$.
 
We have no external information on the possible line flux in any one object, except that $A$ is bound by $[0,1]$. Therefore $p_{prior}(A)$ is flat in [0,1], normalized by:
\begin{equation}
\int^{\infty}_{-\infty}p_{prior}(A) dA = \int^{1}_{0}p_{prior}(A) dA  =1.
\end{equation}



The two priors on $A$ and $\Delta x$ can be combined to:
\begin{multline}
p_{prior}(A|\Delta x) = \\
\ \ \ \ (1 - p_{prior}(\Delta x))\times \delta(A=0) + p_{prior}(\Delta x) \times p_{prior}(A),
\label{eqn:prior}
\end{multline}
where $\delta(A=0)$ is the Kronecker delta function.

The question of whether there is any significant line detection in the entire spectrum then boils down to asking whether there
are any parts of the spectrum for which $p(A > 0|\{S\},\Delta x)>p_{threshold}$ is very high, e.g. $p_{threshold} > 0.997$. In practice, we expect such regions in $\Delta x$ 
to have the extent of the PSF or postage stamp size, $x_{max}$. This formalism is based on the assumption of a single line in the spectrum.
If there are two significant lines, e.g. [O~III] and H$\alpha$, then we would expect two disjoint regions in $\Delta x$ with significant $p(A > 0|\{S\},\Delta x)>p_{threshold}$.

If a single position $\Delta x$ meets the threshold, then we simply translate it into a wavelength $\lambda$, a single value of $A$ which can be transformed into a line flux in physical units (also taking into account the normalized wavelength-dependent grism sensitivity), and an uncertainty on that flux given the distribution of $p_{posterior}(A|\{S\},\Delta x)$.  However, given that we are dealing with three-dimensional spectra, a bright emission line in an object that is not a point source produces significant detections of $A$ at positions near the intrinsic $\lambda_{central}$.  Our best estimate of the central line position is the ``significant" pixel responsible for the maximum peak in $p_{posterior}(A|\{S\})$. 


\subsection{Incorporating Photometric Redshift Priors}
\label{sec:photprior}
\begin{figure}
\includegraphics[width=.45\textwidth]{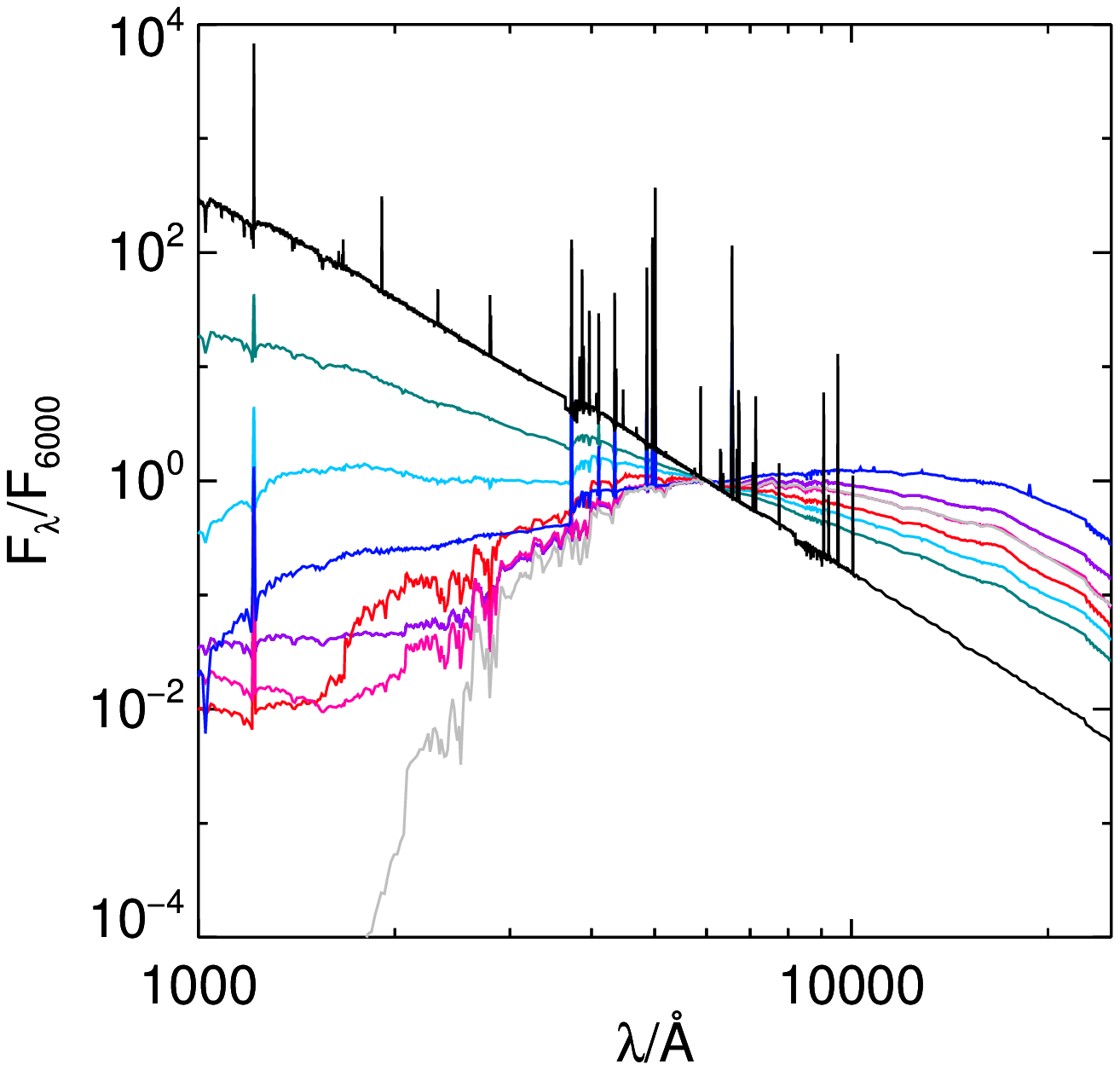}
\caption{Photometric templates used in our application of \textit{EAZY} \citep{eazy} normalized at 6000 \AA.  The colored templates are the default \textit{EAZY} set, created following the \citet{br} algorithm with P\'EGASE models \citep{pegase} and a calibration set of synthetic photometry derived from semi-analytic models, while the black and gray templates are an EELG \cite[from][]{maseda14} and a 1.5 Gyr \citet{bc03} SSP with $A_V=2.5$ to fully reproduce the SEDs of the bluest and reddest objects in the sample.  All galaxy SEDs are fit with linear combinations of these templates.}
\label{fig:temps}
\end{figure}

\begin{figure*}
\begin{center}
\includegraphics[width=.7\textwidth]{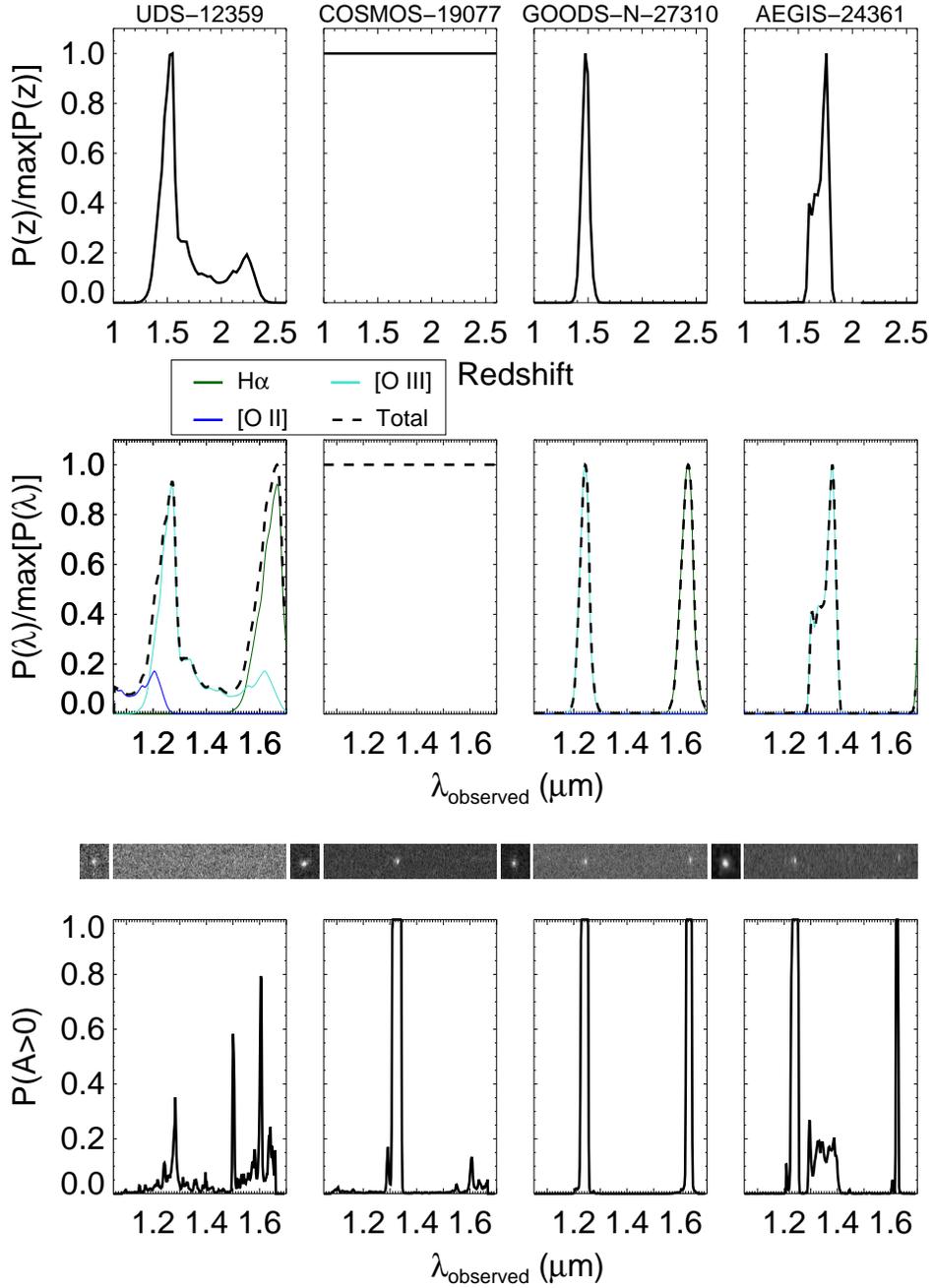}
\end{center}
\caption{Illustration of the line search process for (left to right) \textit{UDS-12359}, \textit{COSMOS-19077}, \textit{GOODS-N-27310}, and \textit{AEGIS-24361}.  From top to bottom: photometric redshift probability distribution from EAZY \cite[$P(z)$;][]{eazy}, the prior on line positions $P(\Delta x)$ derived from $P(z)$ for lines in the correct observed wavelength range, the direct (undispersed) and grism images for the objects, and the output probability at each wavelength position $\Delta x$ that A is nonzero (Equation \ref{eqn:integral}).  The  colored curves denote the expected positions of H$\alpha$, [O III], and [O II] given $P(z)$, while the black curve denotes the overall $P(\lambda)$ for all emission lines that could fall in the grism coverage.  Note that in this case Mg II, [S III], He I, and Pa$\beta$ do not appreciably contribute any probability for these objects in this observed wavelength range.  In the case of  \textit{UDS-12359}, no significant ($> 3\sigma$) line detections are found; for \textit{COSMOS-19077}, \cite[one of the objects studied in][]{maseda13,maseda14} a strong line is discovered despite assuming a flat $P(\lambda)$ due to a high-$\chi^2$ EAZY fit: this object's redshift cannot be reliably determined and is therefore excluded; for \textit{GOODS-N-27310}, the $P(z)$ correctly predicts the positions of the emission lines; and for \textit{AEGIS-24361}, the lines are slightly offset from the predicted position (although we detect them regardless).}
\label{fig:photprior}
\end{figure*}

A strength of this approach is that independent information about the likely redshift of the objects can be folded-in straightforwardly and stringently: it simply gets incorporated as a non-constant $p_{prior}(\Delta x)$ in Equation \ref{eqn:prior}.  Given the amount of ancillary photometry in the 3D-HST/CANDELS fields, spanning from UV to IR wavelengths, specifically 0.3 - 8 $\mu$m, it is straightforward to estimate the redshifts photometrically for the sample \cite[a full description is given in][]{gb}.  Briefly, we calculate photometric redshifts by applying \textit{EAZY}, which calculates model fluxes by convolving linear combinations of high-resolution spectral templates with the filter transmission curves, to the broadband spectral energy distributions (SEDs) in order to estimate a probability distribution for the redshift, $P(z)$.  In addition to the default \textit{EAZY} template sets, we include an additional dusty starforming template \cite[a][SSP of 1.5 Gyr and $A_V=2.5$]{bc03} and an EELG template \cite[the highest sSFR galaxy from][UDS-6195]{maseda14}, as shown in Figure \ref{fig:temps}. We choose to use this $P(z)$ distribution in cases when the minimum reduced $\chi^2$ value is less than 5, which happens in $\sim$90$\%$ of the cases, otherwise we adopt a flat $P(z)$ prior.  This is illustrated in the upper two panels of Figure \ref{fig:photprior}.

For a single line detection in a given spectrum, we do not know which restframe emission feature it corresponds to.  The strongest (blended) emission line complexes we expect to typically observe are Pa$\beta~\lambda$12820, He I $\lambda$10830, [S III] $\lambda$9530, H$\alpha ~\lambda$6563, [O III] $\lambda\lambda$5007,4861, [O II] $\lambda\lambda$3727,3729, and Mg II $\lambda$2800.  This implies that we are only searching for sources with $z \lesssim 5.1$\footnote{This is the redshift where we lose Mg II from the grism coverage.  For the photometric redshifts, \textit{EAZY} is run with $z < 6$.  A search for higher redshift restframe-UV emission lines in this grism data is the subject of ongoing work.}.  We convolve the redshift prior $P(z)$ with the restframe wavelengths of these emission lines to determine a prior probability of a line detection as a function of observed wavelength.  At each wavelength within the G141 grism coverage, we determine the probability of a line being present at the pixel position corresponding that wavelength $P(\Delta x)$ as the combined value of the individual line PDFs at that position.  Examples are shown in Figure \ref{fig:photprior}.  As noted in \citet{skelton}, there are indeed some cases in which the photometric redshift for a given object varies greatly from its spectroscopic redshift.  This (small) percentage varies from field to field and is likely a function of magnitude, so we adopt a floor in our PDF such that only 98$\%$ of the total probability is allocated according to the photometric prior and distribute the remaining 2$\%$ uniformly across all observed wavelengths.

\begin{figure}
\includegraphics[width=.45\textwidth]{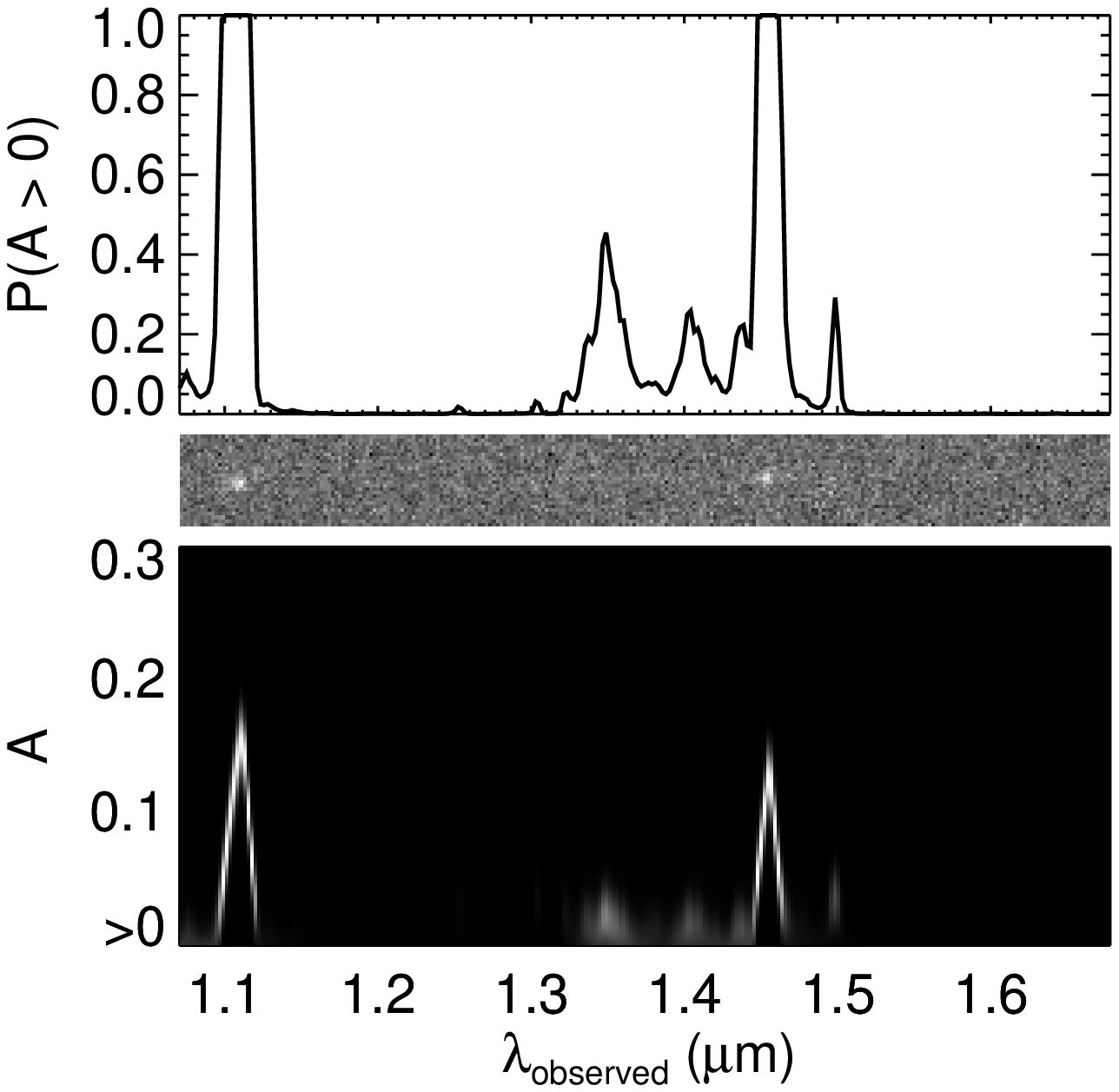}
\caption{Illustration of Equation \ref{eqn:integral} for \textit{GOODS-S-38590}.  The bottom panel illustrates the two-dimensional probability array $p_{posterior}(A|\{S\},\Delta x)$ with the color coding ranging from black (zero probability) to white (high probability).  By marginalizing this probability distribution over all nonzero values of $A$, we arrive at the top panel, which is the same as the bottom set of panels in Figure \ref{fig:photprior}.  The object's $G141$ spectrum is shown in the middle panel.}
\label{fig:photprior2}
\end{figure}

\subsection{Redshifts}
\label{sec:z}


For every measured emission line, we must determine the redshift of the galaxy.  We iteratively assume that the strongest detected line in the grism is Pa$\beta$, He I, [S III], H$\alpha$, [O III], [O II], and Mg II and calculate the detection significance at the predicted positions of the other emission lines.  If we have a significant detection(s) at the predicted position(s), then we have a secure redshift determination.  In the case where we do not find a significant additional emission line, we automatically identify the detected line according to the highest probability for a given line species at that wavelength position according to the photometric redshift information.  

As a final check, we visually inspect all detected lines to verify that the detection is not caused by severe contamination or artifacts at the very edges of the detector\footnote{We do not intend to visually verify the existence of the line, but rather to eliminate cases of obvious contamination.  Our Bayesian framework eliminates the need for subjective visual searches.}: this occurs in less than $\sim$5\% of spectra, see Section \ref{sec:contam}.  We also  classify objects based on the agreement between the emission line redshift and the photometric redshift: some objects have significant detections of lines and photometric redshift information that either does not provide strong constraints at the position of the line and thus the line identification is somewhat dubious or provides no information due to the reduced-$\chi^2$ cut.  Further details on these ``unknown" objects are presented in Section \ref{sec:stack}.



\section{Completeness of the Sample}
\label{sec:completeness}
While grism spectroscopy allows us in principle to search for emission lines in an unbiased manner, several important issues affect our search completeness.  

\subsection{Line Detection Limits}
\label{sec:fluxlimit}
The primary test of the method's efficacy is to insert fake emission lines into spectra and attempt to recover them.  In order to do this, we identify a control sample of 1,425 objects representing a variety of galaxy sizes and morphologies where our search method does not return any spectral positions with a significant detection.  We insert a fake emission line at 1.4 $\mu m$, which is simply the direct image of the object scaled to a given flux value.  We then run our search algorithm, focusing on a $\pm$ 7 pixel region around 1.4 $\mu m$ (corresponding to the average physical extent of a galaxy in this sample based on the Kron radii from the \citeauthor{skelton} \citeyear{skelton} photometric catalog) to see how many lines are recovered as a function of the scaled flux value.  In addition, previous work from 3D-HST has shown that typical starforming galaxies have star formation (as traced by H$\alpha$ emission) out to $\sim 30\%$ larger radii than the rest-frame $R$-band stellar continuum \citep{nelson}.  We also repeat this test making the artificial emission line $30 \%$ larger at the same integrated flux value.

\begin{figure}
\includegraphics[width=.48\textwidth]{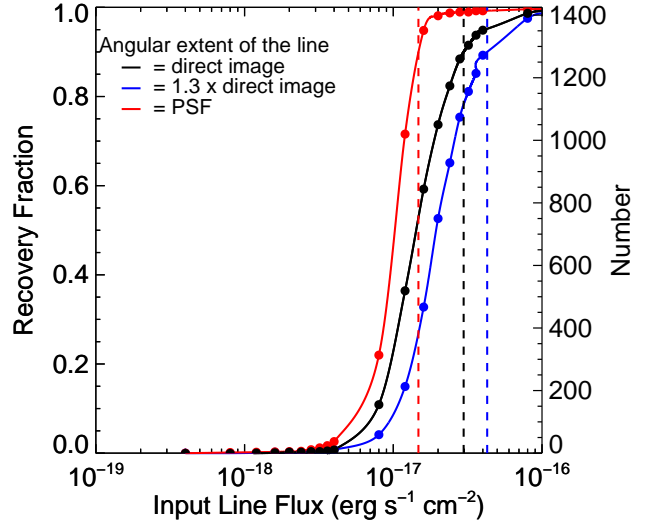}
\caption{Completeness of line recovery test as a function of the fake emission line flux, inserted at 1.4 $\mu m$.  Black denotes an emission line with the same profile as the direct image, blue  denote an emission line that is 1.3 times larger \citep{nelson}, and red denotes an emission line the size of the F140W PSF.  At 90\% completeness, we find flux limits of 3.0$\times$10$^{-17}~\cgs$, 4.4$\times$10$^{-17}~\cgs$, and 1.5$\times$10$^{-17}~\cgs$, with vertical dashed lines denoting these limits.}
\label{fig:linecompleteness}
\end{figure}

The results of this test are shown in Figure \ref{fig:linecompleteness}.   At $90 \%$ completeness, we find a flux limit of 3.0$\times$10$^{-17}~\cgs$ for the compact line case and 4.4$\times$10$^{-17}~\cgs$ for the extended line case.  When we insert an artificial line with the spatial extent of the F140W PSF we obtain a 90\% completeness limit of 1.5$\times$10$^{-17}~\cgs$.  This value is comparable to the theoretical point-source calculation from simulated 3D-HST G141 spectra of 1.6$\times$10$^{-17}~\cgs$ at the same completeness \citep{gb}.  Background-limited grism line searches such as these are primarily sensitive to surface brightness which is why the black curve in Figure \ref{fig:linecompleteness}, which represents a sample with a larger dispersion in object sizes, rises more slowly than the red PSF curve.  When we enlarge \textit{all} emission lines by $30 \%$, we decrease the surface brightness of all galaxies at a given flux and hence we become less sensitive.

\begin{figure}
\includegraphics[width=.42\textwidth]{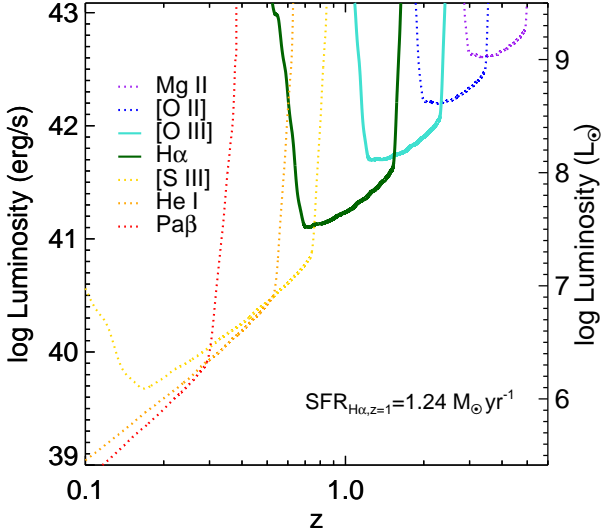}
\caption{Line luminosity at our 90\% completeness limit for direct images, 3.0$\times10^{-17}~\cgs$, as a function of line species and redshift (dotted lines denote the luminosities for lines we do not consider in constructing our ``high-EW" sample; see Section \ref{sec:results}).  The H$\alpha$ star formation rate comes from the calibrations of \citet{k98}.}
\label{fig:sens}
\end{figure}

Line sensitivity will also vary by wavelength, according to the throughput of the grism.  We have performed all of these tests at 1.4 $\mu m$, close to the center of the G141 grism.  The true sensitivity of our method at a given wavelength, then, is the above-quoted line sensitivity scaled according the ratio of the 1.4 $\mu m$ throughput to the throughput at the observed wavelength, see Figure \ref{fig:sens}.  We can convert this flux completeness limit as a function of observed wavelength into a luminosity completeness limit as a function of redshift and line species, as illustrated in Figure \ref{fig:sens}.

\subsection{False Positives}
\label{sec:false}

While the above test determines the flux limit at which an emission line is likely to be recovered, it does not inform us how often a noise peak or artifact would be detected significantly.  In order to investigate this, we isolate contiguous regions of the grism pointing that do not have continuum model fluxes above $5\times10^{-5}~e^-~s^{-1}~px^{-1}$ for each field and create artificial 2D extractions, each 284$\times$31 pixels in size.  As these regions are unlikely to contain real spectral information, any peaks represent noise, unmodeled contamination, or detector artifacts.  

\begin{figure}[t]
\centering
\includegraphics[width=.42\textwidth]{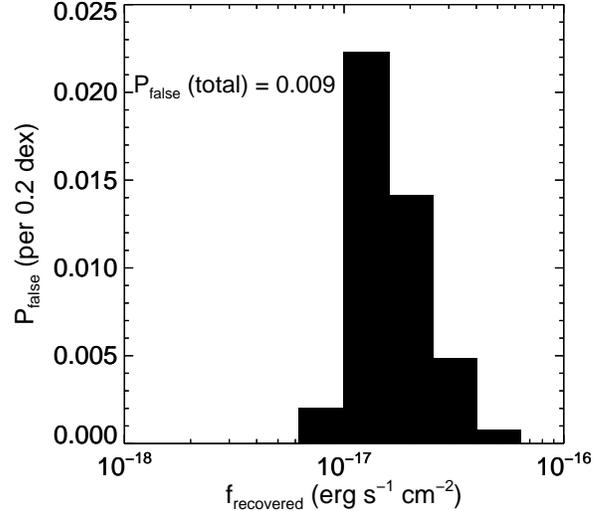}
\caption{Histogram of recovered line fluxes for ``blank'' spectra as part of the search for false positives.  In 0.2 dex wide bins of recovered line flux, we plot the probability that a line with a measured flux is due to a spurious feature.  For the full range in fluxes, redshifts, and equivalent widths, this corresponds to 0.88\%.  The lack of false detections below $10^{-17}~\cgs$ is due to the additional line completeness limits shown in Figure \ref{fig:linecompleteness}.}  
\label{fig:falsepos}
\end{figure}

We create 176 spectra in this manner spread across all fields.  To mimic our standard line search as closely as possible, we randomly assign one of these ``blank" spectra to each of the 159,536 unique objects in the photometric catalog with $m_{det} > 24$ that lie in the 3D-HST spectroscopic footprint and perform the standard cross-correlation analysis, also using its photometric redshift prior.   Throughout, we utilize the \texttt{MAG\_AUTO} value as described in \citet{Momcheva}. This is important when determining the line fluxes as this is the magnitude of the object in the same wavelength range as the grism spectrum and within the same segmentation map which we use as the kernel.  Overall, 1,408 (0.88\%) yielded at least one $>3\sigma$ detection (see Figure \ref{fig:falsepos}) for all line species and line fluxes.


The number of false positives varies as a function of line flux, as not all cosmetic features and noise peaks are of the same magnitude.  As this number is higher than for purely Gaussian noise (which would correspond to 0.27\% for 3$\sigma$ detections), we conclude that the grism exposures contain significant amounts of correlated noise and artifacts that mimic emission features, also due to un-modeled or under-predicted spectral contamination.  Additional criteria are applied to create useful samples (e.g. a cut on EW as in Section \ref{sec:results}), and thus true contamination levels are likely lower than this.

\subsection{Contamination}
\label{sec:contam}
Due to the slitless nature of grism spectroscopy, some sources are strongly contaminated by overlapping spectra from brighter sources.  Chance alignment of sources in the direct image could result in both having ``detections" of the same emission line in the grism data, especially if both sources are spatially small.  There is no automated way to account for such events, so we must resort to visual inspection: for all objects with detected emission lines, we search for all other objects with detected emission lines that lie in a rectangular aperture with an extent corresponding to the $G141$ dispersion size (284 interlaced pixels).  If multiple sources ``produce" the same emission line, we assign the line to a single source based on the overlap with the expected trace of the source (a direct overlap as opposed to a glancing one) and the $F140W$ morphology of the sources.

As described in Section \ref{sec:data} and \citet{Momcheva}, we have a sophisticated flux model for each object in a pointing.  The modeled flux for neighboring sources is subtracted when searching for emission lines in an object's spectrum to avoid potential false detections.  Each object's own flux distribution is also subtracted: the flat flux distribution represents the continuum level of an object, which needs to be subtracted in order to discover residual emission lines.  

The model, however, occasionally does not subtract perfectly and we are left with residual flux.  This typically scales with the flux of the object, such that brighter regions tend to have larger residuals.  Regions of positive residual ($S' - $\textit{Model}) appear like spectral features in that they are areas of ``real" flux above the background level.  These regions are identified as emission features, both in the (bright) object that created the original spectrum and in spectra of objects that happen to overlap with them. We mask out any pixels which have a flux level in the model higher than a threshold value. We seek to strike a balance between masking as few pixels as possible, maximizing our search area, and minimizing the chance of contamination leading to false detections.  We select this masking level by utilizing the same framework as in Section \ref{sec:false}.  We determine the number of line detections at or below $10^{-17}~\cgs$, where we should be only $\sim$10\% complete, as a function of the masking threshold.  The number of ``detections" per pixel is nearly zero when we mask regions where the model flux exceeds 0.004 $e^-~s^{-1}~px^{-1}$.  This level corresponds to the counts in the central pixel from a dispersed point source of an unresolved emission line with a flux of $\sim1.7\times10^{-20}~\cgs$ at 1.4 $\mu$m.  However, once we change to a lower threshold (i.e. the masked region corresponds to brighter fluxes and hence covers a smaller area) we begin to see increasing number counts.  We therefore select this as our masking threshold as it maximizes the usable area while still removing as many potentially problematic regions of the grism frame as possible.



The primary issue for contamination from overlapping spectra, then, comes from the limited area in which we search for lines after applying this masking.  The total unusable area depends on the specific pointing in question, but is equal to 18\% when averaged over the whole survey area with a standard deviation of 4.1\%.  There are some specific cases in which the model fails to account for a particularly bright spectrum, typically in higher-orders for bright stars, and we are left with ``uncontaminated" regions of residual flux.  These cases are obvious to identify and are a reason why all objects with detections are visually inspected.

\subsection{Completeness of the Photometric Catalog}
\label{sec:phot}

This starting point of this search is the photometric catalog of \citet{skelton}.  Therefore we do not analyze spectra for sources that are not in the input photometric catalog.

\begin{figure*}
\includegraphics[width=.95\textwidth]{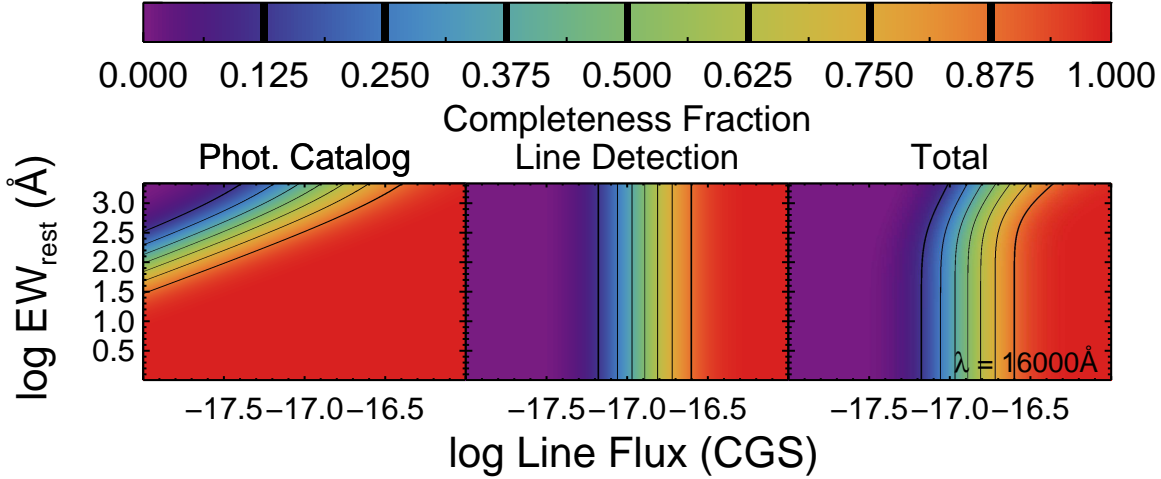}
\caption{Completeness fraction as a function of line flux and equivalent width at a fixed line position (16000~\AA).  The left panel shows the completeness of the \citet{skelton} photometric catalog for objects of a given flux and equivalent width (i.e. continuum magnitude), the central panel shows the flux completeness of our line search from Figure \ref{fig:linecompleteness} and the $G141$ sensitivity at that position, and the right panel shows the combination of these two completeness functions.  This wavelength correspond to H$\alpha$ at $z = 1.4$ or [O III] at $z = 2.2$.  There are small variations in the grism sensitivity as a function of wavelength as well as catalog completeness limits that vary by field, but this information is also taken into account in this analysis.}
\label{fig:totalcompleteness}
\end{figure*}

In the left panel of Figure \ref{fig:totalcompleteness}, we show the completeness fraction of the photometric catalog as a function of line flux and equivalent width, assuming a single emission line is placed in the $H_{F160W}$ filter.  This is the emission line version of Figure 14 in \citet{skelton}.  The $90\%$ catalog completeness limit is $H_{F160W}=25.1$, which corresponds to an emission line flux of $\sim10^{-16}~\cgs$ if entirely concentrated in a single line of infinite equivalent width.  Note that for a given line flux we are more likely to have the object in the photometric catalog if it has a \textit{lower} equivalent width, as that implies the continuum level is higher.

The requirement that an object must be in the photometric catalog is the single strongest prior we apply to our data.  If an emission line source is not in that catalog, by definition we will not be able to detect the line.  In the range of fluxes where we can still robustly detect lines, our catalog completeness is approximately 60\% for high-EW sources.  A full catalog of emission lines for 3D-HST galaxies will be presented in Maseda et al. (in prep.), utilizing a deeper photometric catalog to overcome the issues mentioned here.  A comparison of this method to photometric methods for finding high-EW galaxies, namely the $iJH$-selection of \citet{vdw}, is given in Appendix \ref{sec:appendixa}.  



\section{A SAMPLE OF HIGH-EQUIVALENT WIDTH [O III] AND H$\alpha$ EMITTERS}
\label{sec:results}
We apply our method to 93,832 unique objects in 3D-HST with full spectral coverage.  For objects with multiple spectra due to the overlapping individual pointings, we adopt the redshift corresponding to the highest individual line detection probability.  In total, we find 22,786 objects with at least one emission line.  In order to estimate line equivalent widths, we use the \citet{skelton} catalog $F125W$, $F140W$, or $F160W$ magnitude (depending on the line position) and the measured line flux from \ref{sec:formalism}.

\begin{figure}
\includegraphics[width=.48\textwidth]{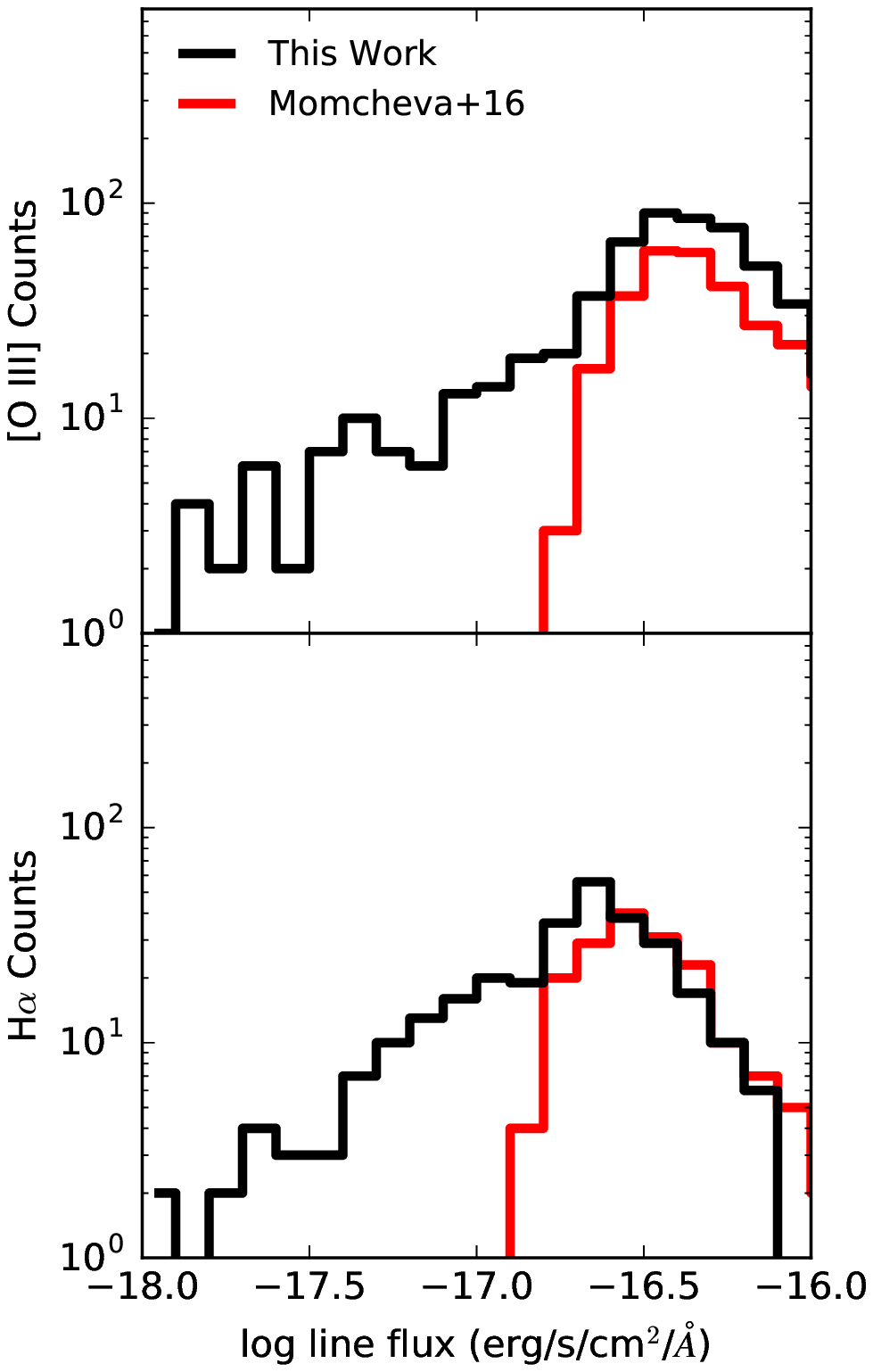}
\caption{Flux comparison between detected emission lines in the method presented here and that in the 3D-HST catalog of \citet{Momcheva}.  The top panel shows the distribution of [O III] emitters and the bottom panel shows the distribution of H$\alpha$ emitters.  In both cases we restrict to objects with 3-$\sigma$ line detections (flux and EW) and continuum $F140W$ magnitudes fainter than 25.  The majority of the lines from this method below fluxes of 10$^{-17}~\cgs$ are measured due to the presence of a stronger line present in the spectrum (see Section \ref{sec:z}).}
\label{fig:linecomp}
\end{figure}

In Figure \ref{fig:linecomp} we show the flux histograms for [O III] and H$\alpha$ emitters from this method and from that of \citet{Momcheva}.  These methods are quite complimentary given the optimizations for faint lines in objects without strong continuum flux presented in this work.

Here, we present a first application of our approach: to measure the number density evolution of extreme EW galaxies with redshift, specifically from 0.7 $< z <$2.3 with an extreme restframe-optical emission line EW ([O III] in excess of 500~\AA~and/or H$\alpha$ in excess of  424 \AA, see Section \ref{sec:stack} for specific information about the cuts used). A full analysis of all line detections and an accompanying catalog is beyond the scope of this methodology paper. With the method outlined in the previous sections we now have, for the first time, a spectroscopic sample of such objects with a well-quantified selection function.  Overall, we have  442 confirmed high-EW [O III] emitters (146 of which have multiple lines) and 340 H$\alpha$ emitters (117 of which have multiple lines). 


Objects with detected lines are put into three classes: multiple significant line detections, single significant line detections with well-known line identifications and hence redshifts from the photometric redshift information, and single significant line detections that have either ambiguous identifications either due to broad photometric priors, namely $P([O~III]) \sim P(H\alpha)$, or because they are detected far away from the expected position from the photometry, like a more extreme case than \textit{AEGIS-24361} in the right panels of Figure \ref{fig:photprior}.  

\subsection{Verification of Redshifts}

\begin{figure}
\includegraphics[width=.45\textwidth]{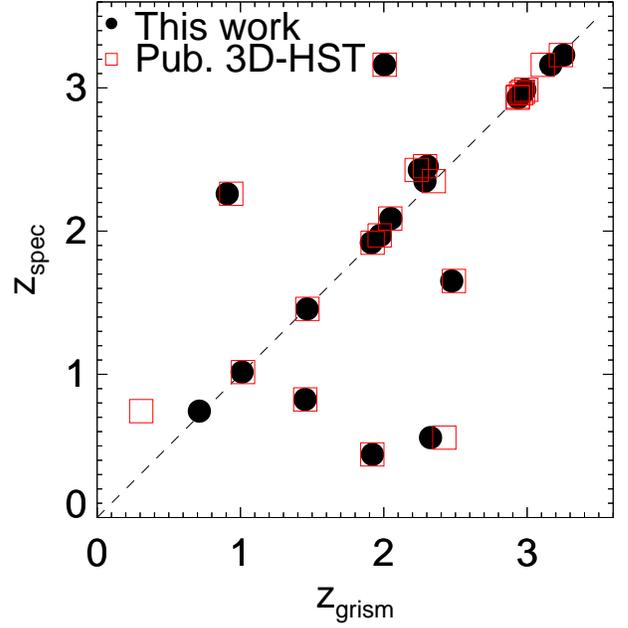}
\caption{Comparison between grism-based redshifts \cite[black circles show results using the method presented in this work, while red squares show results published in][]{Momcheva} and ground-based spectroscopic redshifts in GOODS-N \cite[as presented in][]{skelton}.  There is overall good agreement between the two grism-based methods; some disagreements exist with the published spectroscopic redshifts, but we note that those redshifts are sometimes inaccurate, as shown in Figure \ref{fig:zspec}.}
\label{fig:zcomps}
\end{figure}

In order to demonstrate the accuracy of our new grism redshifts, we compare to existing ground-based spectroscopic redshifts.  Such samples are typically derived at optical wavelengths for the brighter objects in the sample.  While none of our high-EW [O III] and H$\alpha$ emitters have existing published ground-based spectroscopic redshifts, 22 galaxies in our GOODS-N sample have spectroscopic redshifts \cite[see $z_{spec}$ in][]{skelton} as well as robust grism redshifts.  Of these 22, eight have values that disagree by more than $\Delta z = $ 0.1.  We note that all of those eight have published grism redshifts from \citet{Momcheva} that agree with our grism redshifts to better than $\Delta z = $ 0.03 using an independent analysis the same data with a different method.  These values are summarized in Table \ref{tab:zspec} and Figure \ref{fig:zcomps}.

\begin{deluxetable}{lccc}
\tablecaption{Grism and Spectroscopic Redshifts in GOODS-N}
\tablehead{ \colhead{ID} & \colhead{$z_{grism}$} & \colhead{$z_{spec}$} & \colhead{$z_{3D\textendash HST}$}}
\startdata
   5878 & 2.326 & 0.5583 & 2.422\\
  7163 & 2.937 & 2.931 & 2.934\\
  7243 & 1.451 & 0.8250 & 1.452\\
  8614 & 2.287 & 2.349 & 2.349\\
  10728 & 2.975 & 2.973 & 2.971\\
  11429 & 0.9076 & 2.261 & 0.9357\\
  11683 & 1.011 & 1.016 & 1.018\\
  12851 & 2.049 & 2.088 & 2.045\\
  13286 & 2.004 & 3.162 & 2.006\\
  14300 & 1.465 & 1.457 & 1.467\\
  14475 & 2.941 & 2.939 & 2.936\\
  16354 & 2.474 & 1.652 & 2.489\\
  16755 & 1.913 & 1.919 & 1.922\\
  17709 & 3.165 & 3.161 & 3.109\\
  19350 & 2.248 & 2.427 & 2.229\\
  20899 & 2.984 & 2.987 & 2.994\\
  21256 & 2.965 & 2.962 & 2.958\\
  21267 & 1.921 & 0.4410 & 1.920\\
  23343 & 0.7131 & 0.7431 & 0.3076\\
  23744 & 2.303 & 2.453 & 2.288\\
  28202 & 3.255 & 3.229 & 3.235\\
  32925 & 1.976 & 1.970 & 1.971
\enddata
\label{tab:zspec}
\tablecomments{ID numbers and $z_{spec}$ values come from \citet{skelton}; $z_{3D\textendash HST}$ values come from \citet{Momcheva}; $z_{grism}$ values are from this work.}
\end{deluxetable}

\begin{figure}
\includegraphics[width=.48\textwidth]{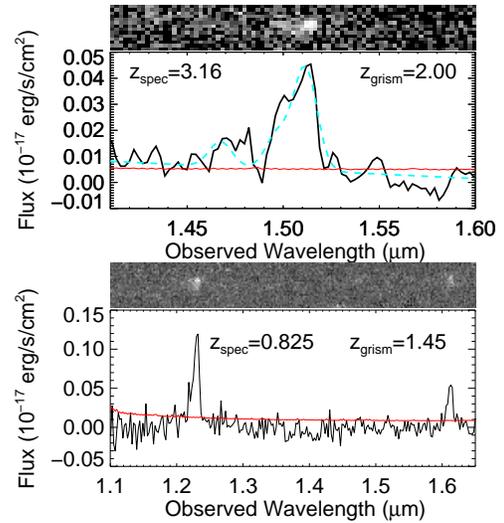}
\caption{Example 3D-HST grism spectra for objects with a disagreement between $z_{spec}$ and $z_{grism}$.  The lower panels show 1D spectra, where black lines are the measured spectral fluxes and red lines are the 1-$\sigma$ errors on the fluxes.  The upper panels show the 2D grism spectra.  (Top) \textit{GOODS-N-13286}, showing an emission line at $\sim1.51~\mu$m.  The published $z_{spec}$ value of 3.16 does not predict an emission line at this position.  We believe this is a clear case of [O III] and H$\beta$ emission due to the clear asymmetry in the bright line (the unresolved [O III] doublet): the cyan curve shows the best-fit [O III]+H$\beta$ model. (Bottom) \textit{GOODS-N-07243}, showing clear emission lines at $\sim1.23$ and $\sim1.61~\mu$m.  The published $z_{spec}$ value of 0.825 does not predict emission lines at these locations, which are consistent with [O III] and H$\alpha$ at $z = $1.45.  In both cases, we suggest that the published spectroscopic redshifts are incorrect.}
\label{fig:zspec}
\end{figure}

When considering the $z_{spec}$ values, some care must be taken.  \citet{skelton} note that, for GOODS-N, ``no quality flags were provided, so there is a mix of reliable and less reliable redshifts in this field.''  Two examples where $z_{spec}$ and $z_{grism}$ disagree are shown in Figure \ref{fig:zspec}.  In these cases we identify a secure redshift based on the combination of the photometric information and the clear asymmetric profile of the unresolved [O III] doublet (upper panel; \textit{GOODS-N-13286}) as well as detections of both [O III] and H$\alpha$ (lower panel; \textit{GOODS-N-07243}).

The redshifts at which this work is most efficient, particularly at $1.1 \lesssim z \lesssim 2.3$ when [O III] is visible in the G141 grism, are generally difficult to confirm from the ground at optical wavelengths.  Comparison to a small subset of objects that have both grism redshifts and ground-based spectroscopic redshifts shows this difficulty.  Further independent verification of the grism-derived redshifts would require additional observations at near-infrared wavelengths.  While large samples of such redshifts are currently being obtained, the multiplexing capabilities of slitless grism spectroscopy using WFC3 are difficult to match.

\subsection{Stacked Spectra}
\label{sec:stack}

We need to determine the fraction of objects in this third category that we can positively identify as H$\alpha$ or [O III] emitters since uncertain or missing photometry can cause problems in the photometric redshift fitting and cannot be properly taken into account as part of our selection function.  Namely, the selection function only contains information from the near-IR imaging, whereas the full galaxy SED is used in the photometric redshift fitting.  While we cannot reliably provide line identifications for individual objects in this category, we can determine the relative numbers of H$\alpha$ and [O III] emitters by comparing spectral stacks with that of known objects.


From high-EW objects with unambiguous redshifts from multiple significant line detections, we would like to determine the typical line ratios.  These ratios are useful to define a flux/EW cut to isolate the same ``extreme" objects, as well as to constrain the relative number of the ``unknown" objects that are respectively [O III] and H$\alpha$.

\begin{figure}
\includegraphics[width=.48\textwidth]{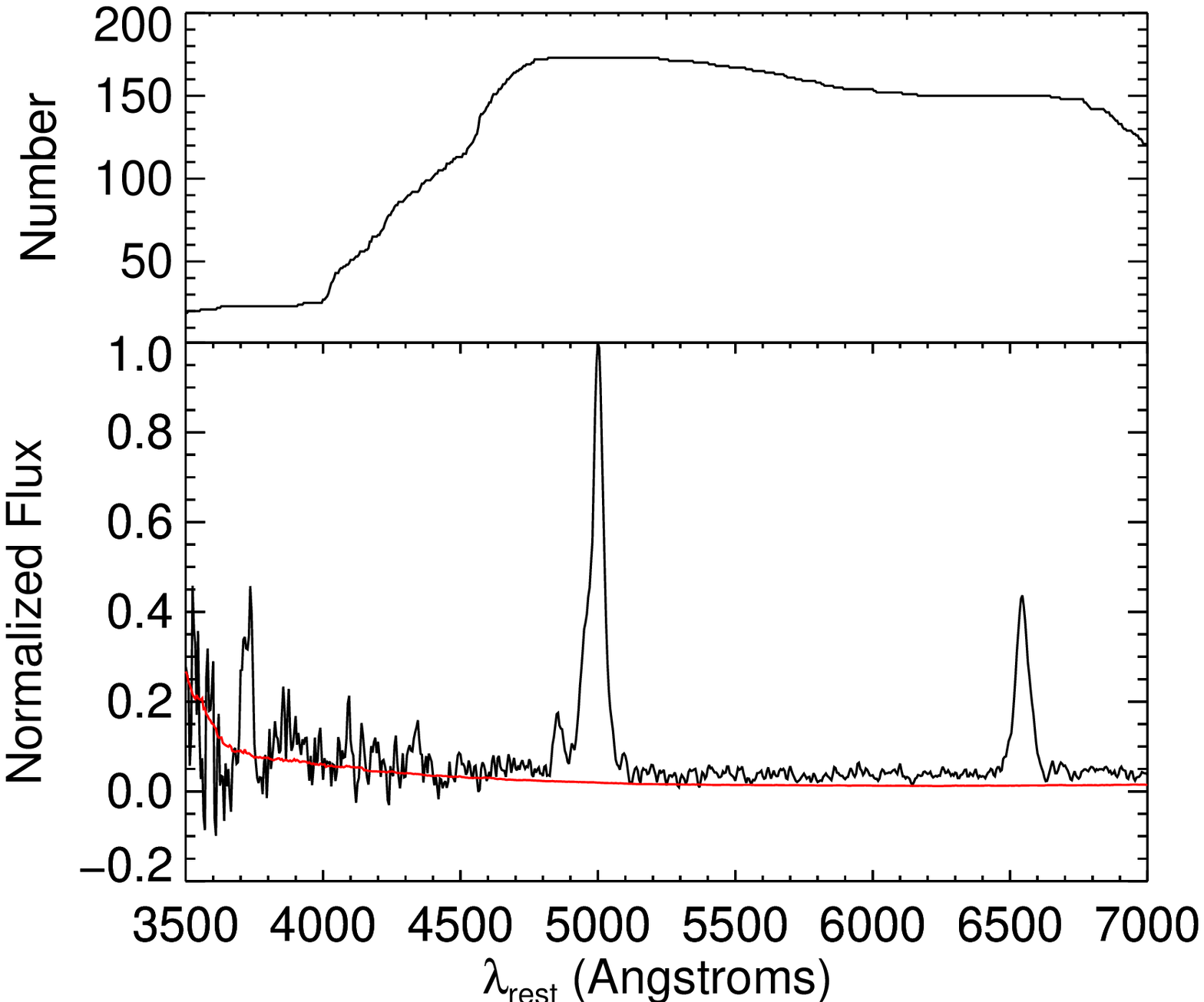}
\caption{Spectral stack for only the objects with known redshifts due to multiple line detections and hence have an unambiguous (photometric) redshift identification, normalized to the [O III] flux.  The upper panel shows the number of unique spectra contributing to the stack at a given wavelength.}
\label{fig:stack}
\end{figure}

To do this, we create a stack of all objects in the first class alone (multiple significant line detections) with restframe $EW_{H\alpha}$ or $EW_{[O III]}$ in excess of 500 \AA, shown in Figure \ref{fig:stack}.  The stacking is done using boxcar extractions of each spectrum in the dispersion direction and weighted according to the grism noise map of the same region.  We consider this spectrum as a template EELG, with line ratios that should be representative of the class.  These objects alone are used since we are primarily interested in the ratio of [O III] to H$\alpha$ and to [O II] and hence need a sample containing objects with multiple lines. The relative line ratios are shown in Table \ref{tab:lines}.

\begin{deluxetable}{lc}
\tablecaption{Stacked Line Ratios\label{tab:lines}}
\tablehead{ \colhead{Lines} & \colhead{Ratio}}
\startdata
$[O III] / H\alpha$&2.03$\pm$0.0327\\
$[O III] / [O II]$&3.50$\pm$0.353\\
$[O III] / H\beta$ &9.50$\pm$0.798
\enddata
\tablecomments{[O III] denotes the single $\lambda$5007 component.  These ratios are the means for the sample as measured where we have spectral coverage of both lines ($1.15 \lesssim z  \lesssim 1.59$ for [O III]/H$\alpha$, $1.88 \lesssim z \lesssim 2.40$ for [O III]/[O II], and $1.21 \lesssim z \lesssim 2.40$ for [O III]/H$\beta$).}
\end{deluxetable}

We use these line ratios to construct an equivalent width limit for the sample.  We define ``high-EW" as having restframe [O III]$\lambda$5007 equivalent width in excess of 500~\AA.  According to the line ratios shown in Table \ref{tab:lines} and a continuum slope $F_{\lambda}\propto\lambda^{-2}$ \citep{vdw}, this implies a restframe H$\alpha$ equivalent width limit of 424 \AA.  

\begin{figure}
\begin{center}
\includegraphics[width=.45\textwidth]{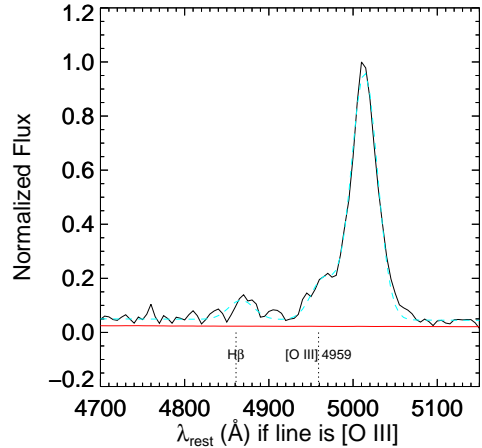}
\end{center}
\caption{Spectral stack for 505 objects with indeterminate redshifts due to single line detections without an unambiguous photometric redshift identification, normalized to the peak line flux.  The $x$-axis shows the wavelengths if the identified emission line is [O III], with the cyan curve showing a Gaussian fit to both [O III] components and H$\beta$ and the red curve showing the combined 1$\sigma$ noise of the stack.  The additional unfit flux blueward of the central line peak can be attributed to [O III]$\lambda$4959 if the line is truly [O III]$\lambda$5007.  The observed ratio of [O III] to H$\beta$ is 140\% of the value we would expect if all objects were [O III] emitters.  As we do not observe a significant detection at the expected position of other lines if the peak is H$\alpha$, we conclude that the majority of these ``unknown" objects are actually [O III] emitters.}
\label{fig:unknownstack}
\end{figure}

We can then create a similar weighted-mean stack, normalized to the peak line flux, for all 505 of the objects in the third category (single line detections that do not have secure identifications and redshifts), shown in Figure \ref{fig:unknownstack}.  Here, we put the detected emission lines at the same wavelength and apply an observed frame equivalent width limit of 1250 \AA~ (equivalent to a restframe EW of 500 \AA~ at $z=1.5$).  We now use this stack to estimate the fraction of [O III] emitters in this stack by correcting the observed ratio of the peak flux (assuming the line is [O III]$\lambda$5007) to the flux at the expected position of H$\beta$ according to the intrinsic ratio from the stack of secure objects.  We do not expect any emission line contribution at this spectral position if the primary line is H$\alpha$ or [O II].  The observed ratio of [O III] to H$\beta$ is 13.3$\pm$1.92, consistent with the observed value within 2-$\sigma$.  Likewise, we can perform the same exercise with the $\lambda$4959 peak of [O III], compared to $\lambda$5007.  This result is 0.50$\pm$0.034 of the expected 3:1 ratio of $\lambda$5007 to $\lambda$4959.  We note that uncertainties in the object centering due to the low spectral resolution and variations in the emission line morphology can reduce the measured 4959-to-5007 ratio, implying that this 50\% could still be a lower limit.

If the primary peak corresponded to H$\alpha$, the same test is somewhat more difficult given that we do not expect to see any other strong emission lines from $0.6 \lesssim z \lesssim 1.1$.  While He I and [S III] are covered to varying degrees in this redshift range, they are typically not strong enough to confirm H$\alpha$ in the absence of other information.  If we assume the lines are H$\alpha$, we do not detect any feature at the expected wavelength of [OIII], placing a 3-$\sigma$ upper-limit on the [OIII]/H$\alpha$ ratio of 0.02.  Compared to the [OIII]/H$\alpha$ ratio from the stack of the objects with secure redshifts, we can conclude that less than 1\% of the objects in this sample are H$\alpha$ emitters.  Note that the expected positions of the [S II] lines ($\lambda\lambda$6717, 6731) are 5126 and 5137 \AA~ on the wavelength scale of Figure \ref{fig:unknownstack} and are not clearly detected, further implying that a significant fraction of these emission lines are not H$\alpha$.

Therefore, our stacked spectra imply that the majority of these objects are [O III] emitters.  We subsequently include these objects as such in the primary sample, putting them at their [O III] redshift and utilizing the [O III] equivalent width selection criteria.

As such, we have 470 ``unknown" objects with single emission lines satisfying the [O III] selection criteria, which we include with the primary [O III] sample of 442 for a total of 912.  Since we can only make the statistical statement that most of these are [OIII] emitters, we must assume that the density estimates for [OIII] emitters in the next section are slightly overestimated.

\subsection{Number Density Evolution}
\label{sec:nd}
\begin{figure*}
\includegraphics[width=.48\textwidth]{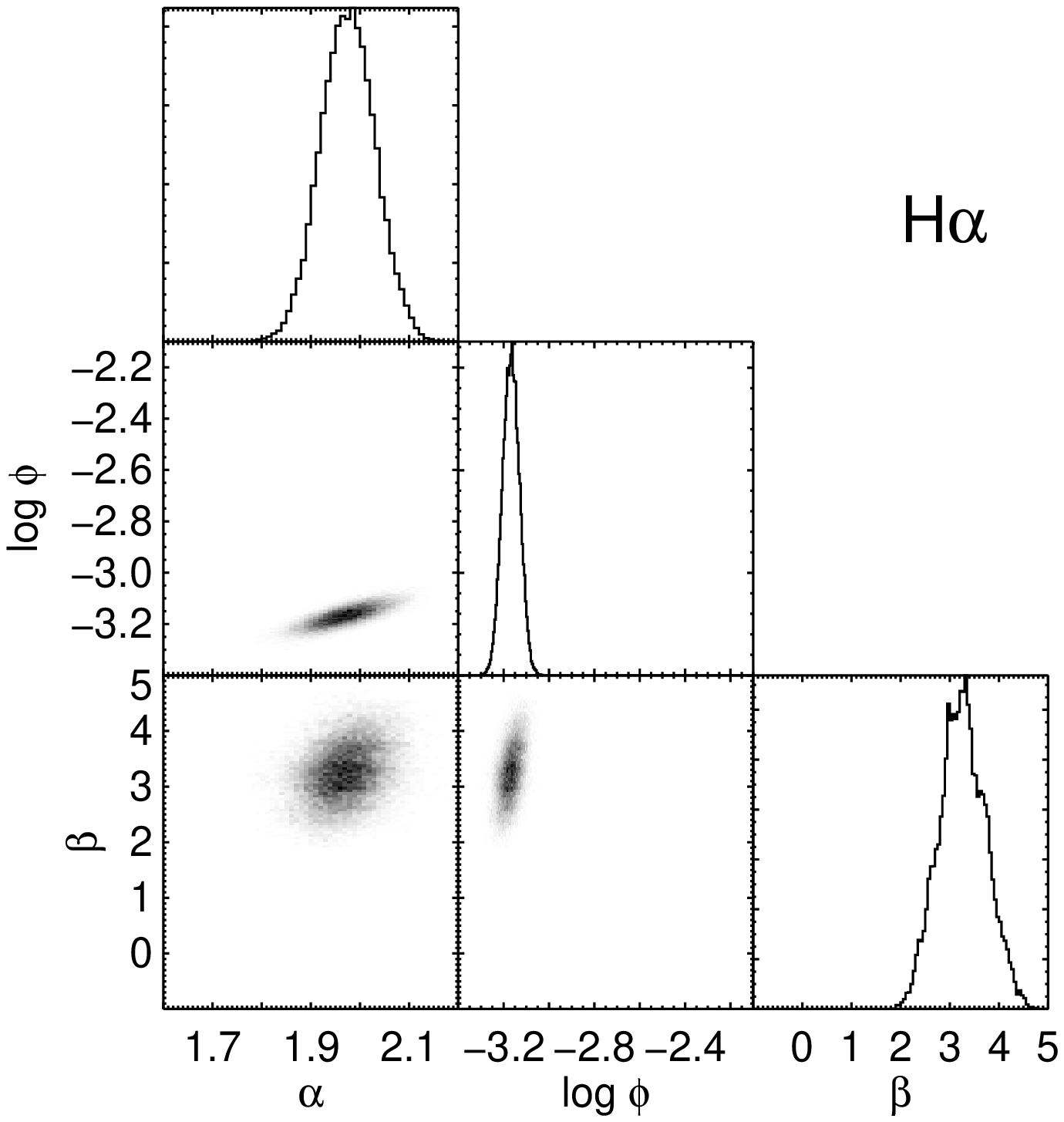}
\includegraphics[width=.48\textwidth]{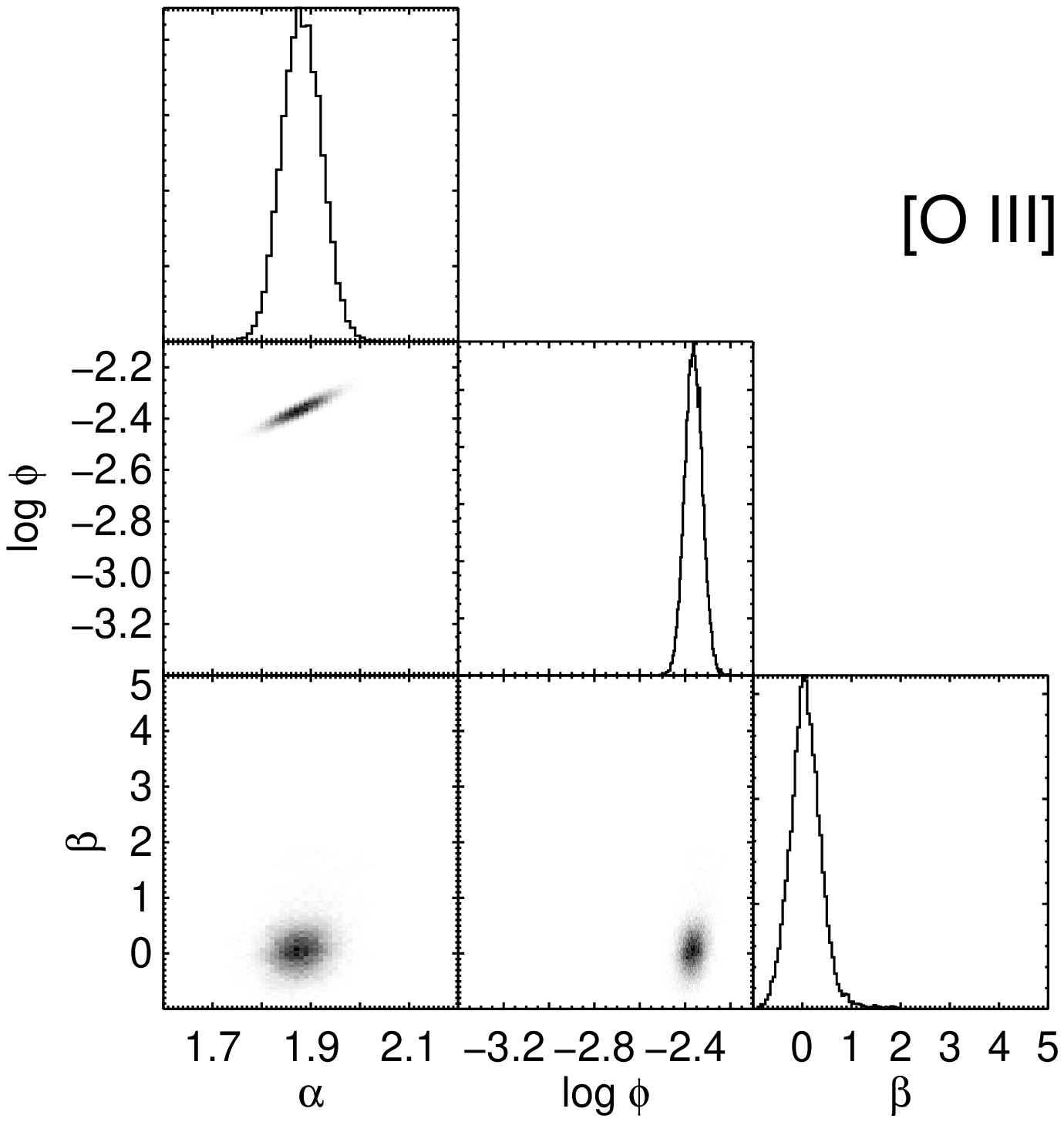}
\caption{Posterior probability distributions of the parameters $\alpha$ and $\beta$ from Equation \ref{eqn:lf} as well as the comoving number density $\phi$ for the H$\alpha$ and [O III] samples.}
\label{fig:corner}
\end{figure*}

To estimate for the evolution in the number density of such high-EW objects, we first need to construct a parameterized functional form.  For simplicity, we assume a power law distribution in line luminosities and a power-law dependence in $(1+z)$.  This functional form, for luminosities in the range $L,~L+dL$ and redshifts in the range $z,~z+dz$, is:

\begin{equation}
\Phi(L,z|\alpha,\beta) = \left(\frac{L}{L_0}\right)^{-\alpha}\left(\frac{1+z}{1+z_0}\right)^{\beta},
\label{eqn:lf}
\end{equation}
where $L_0$ and $z_0$ are fiducial values, here taken to be the median of the [O III] and H$\alpha$ luminosities ($L_0=10^{42}$ erg s$^{-1}$) and the median redshift for the two emission lines ($z_{0,H\alpha}=1.23$ and $z_{0,[O III]}=1.67$).  Details of the calculations are given in Appendix \ref{sec:appendixb}, but in general we can obtain estimates for $\alpha$ and $\beta$ as well as the intrinsic number density of sources in our survey volume, $\phi$ (the normalization, i.e. the total number density of sources that could be observed taking into account the incompleteness per unit volume; see Appendix \ref{sec:appendixb}), per comoving volume element of the survey using a standard Markov Chain Monte Carlo algorithm.  We incorporate the knowledge of our selection function (a numerical combination of the completeness functions described in Sections \ref{sec:fluxlimit} and \ref{sec:phot} and shown in Figure \ref{fig:totalcompleteness}, expressed in terms of line luminosity and redshift) as well.  Results of this analysis are shown in Figure \ref{fig:corner}.

\begin{deluxetable}{lccc}
\tablecaption{MCMC Results\label{tab:mcmc}}
\tablehead{ \colhead{Line} & \colhead{Parameter} & \colhead{Value} & \colhead{Uncertainty}}
\startdata
$[O III]$ & $\alpha$ & 1.88 & $_{-0.0414}~^{+0.0376}$\\
& $\beta$ & 0.0600 & $_{-0.299}~^{+0.296}$\\
& log $\phi~(Mpc^{-3})$ & -2.36 & $_{-0.0434}~^{+0.0326}$\\
$H\alpha$ & $\alpha$ & 1.98 & $_{-0.0608}~^{+0.0482}$\\
& $\beta$ & 3.25 & $_{-0.426}~^{+0.522}$\\
& log $\phi~(Mpc^{-3})$ & -3.17 & $_{-0.0406}~^{+0.0333}$\\
\enddata
\tablecomments{Values are the median of the distributions shown in Figure \ref{fig:corner} for $L_0$ = 10$^{42}$ erg s$^{-1}$ and $z_{0,H\alpha}$ = 1.23 and $z_{0,[OIII]}$ = 1.67.  The uncertainties are the shortest 68\% confidence intervals from these same histograms.  The quoted comoving number density $\phi$ is simply $N$ (the intrinsic number of objects in the field) divided by the volume of the survey in cubic Mpc ($\sim2.7\times10^{6}$ for [O III] and $\sim1.7\times10^{6}$ for H$\alpha$).}
\end{deluxetable}

\begin{figure}
\includegraphics[width=.48\textwidth]{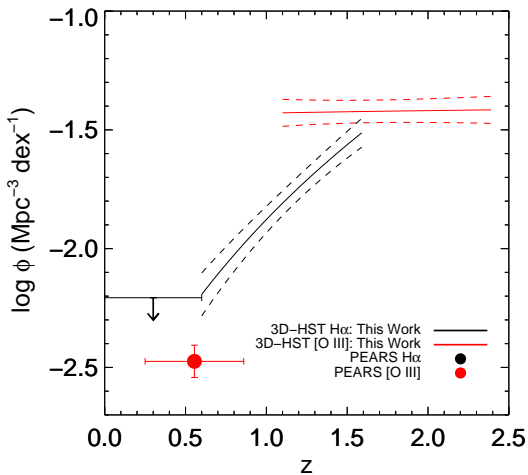}
\caption{Comoving number densities per dex in luminosity, compared at $L=10^{41.5}~erg~s^{-1}$, as a function of redshift for objects with restframe [O III] (red) and/or H$\alpha$ (black) EWs in excess of 424/500 \AA$~$from \textit{PEARS} \citep{pears} and this study.  The dashed lines represent the $\pm$1-$\sigma$ uncertainties in the number densities from the width of the MCMC probability distributions for $\beta$ and $N$.  Error bars in redshift represent the actual redshift range in each bin and not uncertainties in the redshift determination.  As the H$\alpha$ density from \textit{PEARS} is not well-constrained, the upper-limit is plotted.}
\label{fig:numdens}
\end{figure}

After calculating the respective volumes of the survey given the observed area of 723.3 square arcminutes, we can estimate the comoving number density of high-EW [O III] and H$\alpha$ emitters.  These results are shown in Figure \ref{fig:numdens} at a fiducial luminosity of $10^{41.5}~erg~s^{-1}$. We clearly observe a decrease of the comoving number density towards the present epoch from our 3D-HST sample.  Although the distribution for the redshift evolution $\beta$ for the H$\alpha$ sample is wide, the value of $\phi$ is well-constrained from the MCMC analysis and hence the overall uncertainty in the number density evolution, shown by the dashed black lines in Figure \ref{fig:numdens}, is only about 0.1 dex at the low-redshift end.  In general, the individual distributions of $\beta$ in for H$\alpha$ and [O III] in Figure \ref{fig:corner} are consistent with a value greater than zero, implying a secure positive evolution in the comoving number density of high-EW objects.  Across the redshift range probed in this study, this evolution is on the order of $\sim$ 0.8 dex.  Note that this analysis neglects the impact of large-scale clustering, which is expected to be moderate for low-mass systems \citep{boerner}.  Likewise, many authors fit line luminosity functions as Schechter functions, which also include an exponential cutoff at high luminosities.  Since we do not \textit{a priori} know the luminosity function shape at high-EW, we choose the simpler case of a pure power law \cite[see the $0.9 < z < 1.5$ $\lbrack$O II$\rbrack$ sample from][]{pears}.  The narrow distributions for $\alpha$ in both the H$\alpha$ and [O III] samples presented here justify this choice.

Ideally we would have low-redshift samples selected in the same manner as our higher-redshift 3D-HST sample; unfortunately existing large spectroscopic samples of EELGs, such as those of \citet{amorin14a,amorin14b}, are incomplete and do not have well-defined selection functions.  For comparison, we use the results from \textit{PEARS} \citep{pears} with the same equivalent width limits applied to H$\alpha$ and [O III] as in the 3D-HST sample.  While the \textit{PEARS} sample is also derived from grism data, the luminosity limits are sufficiently different from the 3D-HST limits presented here that it is difficult to make a meaningful comparison on an equivalent width-selected sample.  Namely, \textit{PEARS} becomes incomplete at luminosities of $10^{41.5}~erg~s^{-1}$ for [O III] and H$\alpha$.  In their main sample, there are 6 H$\alpha$ lines and 25 [O III] lines that meet our EW limits with a quoted quality value greater than 2.5.  In order to estimate the number density of these objects, we must correct the counts for (flux) incompleteness using their published completeness curves.  As both samples are small, we expect Poisson errors in the number counts to dominate the errors and hence we do not attempt to perform the same MCMC analysis with all parameters free.  Instead, we fix $\beta$ to zero and adopt the $\alpha$ value for H$\alpha$ from the 3D-HST data, fitting only for the normalization.  Figure \ref{fig:numdens} shows consistency between the [OIII] emitters from PEARS and the H$\alpha$ emitters from this work given that we do not fit the PEARS sample with a redshift evolution. In the case of H$\alpha$, the sample is so small that the result is not well-constrained and we only plot the upper-limit.  If we extrapolate our measured H$\alpha$ number density evolution down to the redshifts of the PEARS H$\alpha$ sample and take into account the different survey volumes for these respective redshift ranges (3D-HST = 56 $\times$ PEARS), then we would predict 5 high-EW H$\alpha$ emitters at $L=L_0$ based on the number counts from 3D-HST, modulo differences in the completeness functions for the two surveys.  This expectation value is consistent with the measured counts from the PEARS survey.


We therefore deduce that there is a positive evolution in the number of high-EW objects with redshift, out until at least $z\sim2.4$.

\section{Summary}
\label{sec:summary}
We present here a new method for detecting emission lines in slitless spectroscopic data.  We estimate the likelihood that a given position in the 2D spectrum contains an emission line by cross correlating the position with a kernel corresponding to the direct image and transform this into the probability that the position corresponds to an emission feature.  This method simultaneously includes prior information on line positions from photometric redshift estimates, which also allow us to determine the redshift of the galaxy even when only a single line is detected.  The photometric redshifts are determined using a set of galaxy templates which are appropriate for high-EW objects, which are otherwise poorly fit.  Robust tests of the method using real and simulated data reveal low levels of contamination and yield a well-defined selection function, which depends on the continuum magnitude of the source as well as the emission line wavelength and flux.

Applying this method to the full 3D-HST data set, we obtain a sample of 22,786 galaxies with definite redshifts, either from multiple line detections or single line detections combined with photometric redshift information.  The median line flux of the confirmed objects is $2.7\times10^{-17}~\cgs$.  This method can be considered complimentary to the one presented in \citet{Momcheva} given the relative differences in the complexities of the methods and their abilities to find emission lines in different regimes of continuum strength. These redshifts will be included in a future 3D-HST data release.

Of this sample, we have 782 high-EW [O III] and/or H$\alpha$ emitters, where the EW limits are 500 and 424 \AA, respectively, due to an intrinsic flux difference between the two species in these objects.  Many objects in the survey have detected emission lines but either have a very poorly constrained photometric redshift (i.e. a very high reduced-$\chi^2$ value) or no photometric redshift at all (i.e. it was not detected in enough photometric bands to be fit).  A stacked spectrum for these objects reveal that a majority of them are plausibly [O III] emitters based on the observed [O III]/H$\beta$ line ratio in the stack, and we thus include an additional 470 objects into the [O III] sample.  We therefore have a sample of 912 high-EW [O III] emitters and 340 H$\alpha$ emitters.

We parameterize the number density evolution function, assumed to be a power law in both luminosity and redshift, and probe the posterior probability distributions using a Markov Chain Monte Carlo algorithm.  Our sample shows an increase in the number of these objects with redshift by a factor of 6 from $z\sim0.6-2.4$.  These results strongly suggest a factor of 30 or more evolution between $z\sim2$ and present, even though direct comparisons with lower-redshift studies are difficult due to their small volumes and different line luminosity limits.

The observed positive trend in number density with redshift is implied by previous EELG studies, such as \citet{kakazu} and \cite{vdw} who estimate that [O III] EW $\gtrsim 500$ \AA~galaxies are two orders of magnitude more common at $z\sim1.7$ than at $z=0$, which is in rough agreement with the results found here.  Similarly, \citet{maseda14} argue that their results, combined with precise alignments of the two strong galaxy-galaxy lensing systems discovered in CANDELS/3D-HST \citep{gb2,vdw13}, show that EELGs must be common at $z>1$.  At even higher redshifts, galaxies with these extreme emission line EWs are abundant and perhaps even ubiquitous \cite[at $z>6$; e.g.][]{smit}.

This method, while specifically applied to \textit{HST} grism spectroscopy, is more generally applicable to any spectroscopy with spatial as well as spectral information.  Its automated nature can be utilized to construct samples of emission line galaxies in large surveys, and it can also be used to determine the significance of an emission line ``detection" for individual objects. Planned future grism surveys, such as \textit{Euclid} and \textit{WFIRST}, will require an automated line search to analyzing the huge volume of data produced.  Grisms also feature prominently on the \textit{James Webb Space Telescope}, with both \textit{NIRISS} and \textit{NIRCam} providing multiple slitless grism spectroscopic modes.  Given the variety of photometry that will likely be obtained in parallel, a method such as this one that can combine the spectroscopic and photometric information in a statistical way will be a powerful tool in making best use out of future JWST slitless grism spectroscopic surveys.

\acknowledgements{We would like to thank Greg Rudnick, Fabian Walter, and David Hogg for interesting discussions about statistics.  We would also like to thank Pieter van Dokkum and the rest of the 3D-HST collaboration for their work on the survey and input at various stages of this project.  M.V.M. was a member of the International Max Planck Research School for Astronomy and Cosmic Physics at the University of Heidelberg, IMPRS-HD, Germany.  B.F.L. acknowledges support from the NSF Astronomy and Astrophysics Fellowship grant AST-1202963.}

{\it Facilities:} \facility{HST}.
\bibliographystyle{apj}

\begin{thebibliography}{}
\bibitem[Amor\'in et al.(2014a)]{amorin14a} Amor\'in, R. O., P\'erez-Montero, E., Contini, T., et al. 2014a, arXiv:1403.3441
\bibitem[Amor\'in et al.(2014b)]{amorin14b} Amor\'in, R. O., Sommariva, V., Castellano, M., et al. 2014b, A\&A, 568L, 8
\bibitem[Atek et al.(2010)]{wisp} Atek, H., Malkan, M., McCarthy, P., et al. 2010, ApJ, 723, 104
\bibitem[Atek et al.(2014)]{atek14} Atek, H., Kneib, J.-P., Pacifici, C., et al. 2014, ApJ, 789, 96
\bibitem[Blanton \& Roweis(2007)]{br} Blanton, M. R., \& Roweis, S. 2007, AJ, 133, 734
\bibitem[Boerner, Mo, \& Zhou(1989)]{boerner} Boerner, G., Mo, H., \& Zhou, Y. 1989, A\&A, 221, 191
\bibitem[Brammer, van Dokkum, \& Coppi(2008)]{eazy} Brammer, G. B., van Dokkum, P., \& Coppi, P. 2008, ApJ, 686, 1503
\bibitem[Brammer et al.(2012a)]{gb} Brammer, G. B., van Dokkum, P., Franx, M., et al. 2012a, ApJS, 200, 13
\bibitem[Brammer et al.(2012b)]{gb2} Brammer, G. B., S\'anchez-Janssen, R., Labb\'e, I., et al. 2012b, ApJ, 758, L17
\bibitem[Bruzual \& Charlot(2003)]{bc03} Bruzual, G., \& Charlot, S. 2003, MNRAS, 344, 1000
\bibitem[Cardamone et al.(2009)]{greenpea} Cardamone, C., Schawinski, K., Sarzi, M., et al. 2009, MNRAS, 399, 1191
\bibitem[Cooke(2009)]{cooke} Cooke, J. 2009, ApJ, 704, L62
\bibitem[Ferreras et al.(2012)]{ferreras} Ferreras, I., Pasquali, A., Khochfar, S., et al., 2012, AJ, 144, 47
\bibitem[Fioc \& Rocca-Volmerange(1997)]{pegase} Fioc, M., \& Rocca-Volmerange, B. 1997, A\&A, 326, 950
\bibitem[Grogin et al.(2011)]{candels1} Grogin, N., Kocevski, D., Faber, S., et al. 2011, ApJS, 197, 35
\bibitem[Hathi et al.(2009)]{hathi} Hathi, N. P., Ferreras, I., Pasquali, A., et al. 2009, ApJ, 690, 1866
\bibitem[Kakazu, Cowie, \& Hu(2007)]{kakazu} Kakazu, Y., Cowie, L. L., \& Hu, E. M., 2007, ApJ, 668, 853
\bibitem[Kennicutt(1998)]{k98} Kennicutt, R.~C., Jr.\ 1998, \araa, 36, 189
\bibitem[Koekemoer et al.(2011)]{candels2} Koekemoer, A., Faber, S., Ferguson, H., et al. 2011, ApJS, 197, 36
\bibitem[K\"ummel et al.(2009)]{axe}K\"ummel, M., Walsh, J. R., Pirzkal, N., Kuntschner, H., \& Pasquali, A. 2009, PASP, 121, 59
\bibitem[Lilly et al.(2007)]{lilly} Lilly, S. J., Le F\'evre, O., Renzini, A., et al. 2007, ApJS, 172, 70
\bibitem[Malhotra et al.(2005)]{malhotra} Malhotra, S., Rhoads, J. E., Pirzkal, N., et al. 2005, ApJ, 626, 666
\bibitem[Maseda et al.(2013)]{maseda13} Maseda, M. V., van der Wel, A., da Cunha, E., et al. 2013, ApJ, 778, L22
\bibitem[Maseda et al.(2014)]{maseda14} Maseda, M. V., van der Wel, A., Rix, H. -W., et al. 2014, ApJ, 791, 17
\bibitem[Masters et al.(2012)]{masters12} Masters, D., McCarthy, P., Burgasser, A. J., et al. 2012, ApJ, 752, L14
\bibitem[Meurer et al.(2007)]{meurer} Meurer, G. R., Tsvetanov, Z. I., Gronwall, C., et al. 2007, ApJ, 134, 77
\bibitem[Momcheva et al.(2016)]{Momcheva} Momcheva, I. G., Brammer, G. B., van Dokkum, P., et al. 2016, ApJS, 225, 27
\bibitem[Nelson et al.(2012)]{nelson} Nelson, E. J., van Dokkum, P. G., Brammer, G. B., et al. 2012, ApJ, 747, L28
\bibitem[Patel et al.(2013)]{patel} Patel, S. G., Fumagalli, M., Franx, M., et al. 2013, ApJ, 778, 115
\bibitem[Pirzkal et al.(2004)]{pirzkal} Pirzkal, N., Xu, C., Malhotra, S., et al. 2004, ApJS, 154, 501
\bibitem[Pirzkal et al.(2013)]{pears} Pirzkal, N., Rothberg, B., Ly, C., et al. 2013, ApJ 772, 48
\bibitem[Rhoads et al.(2009)]{rhoads} Rhoads, J. E., Malhotra, S., Pirzkal, N., et al. 2009, ApJ, 697, 942
\bibitem[Schmidt et al.(2013)]{schmidt} Schmidt, K., Rix, H.-W., da Cunha, E., et al. 2013, MNRAS, 432, 285
\bibitem[Skelton et al.(2014)]{skelton} Skelton, R. E., Whitaker, K. E., Momcheva, I. G., et al. 2014, ApJS, 214, 24
\bibitem[Smit et al.(2014)]{smit} Smit, R., Bouwens, R. J., Labb\'e, I., et al. 2014, ApJ, 784, 58
\bibitem[Straughn et al.(2008)]{straughn} Straughn, A. N., Meurer, G., Pirzkal, N., et al. 2008, AJ,  135, 1624
\bibitem[van Dokkum et al.(2011)]{vd} van Dokkum, P. G., Brammer, G. B., Fumagalli, M., et al. 2011, ApJ, 743, L15
\bibitem[van Dokkum et al.(2013)]{vd13} van Dokkum, P. G., Leja, J., Nelson, E. J., et al. 2013, ApJ, 771, L35
\bibitem[van der Wel et al.(2011)]{vdw} van der Wel, A., Straughn, A., Rix, H.-W., et al. 2011, ApJ,  742, 111
\bibitem[van der Wel et al.(2013)]{vdw13} van der Wel, A., van de Ven, G., Maseda, M., et al. 2013, ApJ, 777, L17
\end{thebibliography}

\appendix
\section{Appendix A: FIDELITY OF PHOTOMETRIC SEARCHES}
\label{sec:appendixa}
The photometric selection technique of \citet{vdw} utilizes the $I_{F814W}$-, $J_{F125W}$-, and $H_{F160W}$-bands to preferentially select systems dominated by strong emission lines.  By looking for a flux excess in $J$ compared to the continuum as measured in $I$ and $H$, they claim to select [O III] emitters at 1.6 $< z <$ 1.8, with perhaps minor contamination by H$\alpha$ emitters at z$\sim$1.

As we now have more complete photometry from 3D-HST and CANDELS since the study of \citet{vdw}, we can apply the same photometric cuts to a larger sample.  Previously, \citet{vdw} found 69 objects in 279 arcmin$^2$.  Here, using the same cuts, we discover 312 objects in the full 896 arcmin$^2$ of the 3D-HST survey: 94 in AEGIS, 48 in COSMOS, 67 in GOODS-N (using the F775W filter in place of the F814W filter), 51 in GOODS-S, and 52 in UDS.




We compare this photometric sample with our spectroscopic sample to test the fidelity of the photometric search.   For the following analysis, we ignore objects whenever severe contamination would prevent a line identification and objects without full spectral coverage bluewards of 14000 \AA: we are left then with a total of 186 objects, of which 147 (79\%) have a strong emission line in the $J$-band.  Many of the objects without a detected emission line are intrinsically faint, and thus their emission line could simply be fainter than the noise level in the grism frames.  We thus verify the $iJH$-selection as an efficient way to select emission line galaxies.



Another photometric selection is given in \citet{greenpea} for lower-redshift emission line galaxies, the so-called ``green pea" galaxies.  While the same selections could yield a sizable sample in our data set, we would not detect the strongest emission lines (H$\alpha$ or [O III]) in the NIR for them given their low redshifts.  

\citet{cooke} also develop a selection technique to select emission line galaxies from broadband photometric data, specifically searching for $z\sim3$ Lyman-$\alpha$ emitters at optical wavelengths.  Of a sample of 17 galaxies that were selected using photometric cuts, 8 (47\%) were confirmed to have Lyman-$\alpha$ emission even though the average EW of the emission is a factor of $\sim$ 10 lower than the [O III] emission in the $z\sim1.7$ EELGs.


\section{Appendix B: Bayesian Luminosity Function}
\label{sec:appendixb}
As described in the text, we would like to determine the evolution in the comoving number density of the high-EW emission line sample.  This is a simplified case where all redshifts and luminosities are precisely known, as well as the selection function.

We can parameterize the distribution of sources in the range $L,~L+dL;~z,~z+dz$ as the combination of a power law in luminosity and in redshift according to:
\begin{equation}
\Phi(L,z|\alpha,\beta) = \left(\frac{L}{L_0}\right)^{-\alpha}\left(\frac{1+z}{1+z_0}\right)^{\beta},
\end{equation}
where $L_0$ and $z_0$ are fiducial values (the medians of the input observations).  The probability density of $L$ and $z$, then, is:

\begin{equation}
P(L,z|\alpha,\beta, \phi_0) = \phi~\Phi(L,z|\alpha,\beta)~\frac{dV}{dz}(z),
\end{equation}

where $\phi$ is the total number density of sources N/V; $Mpc^{-3}$) that could be observed and $dV/dz$ is the comoving volume element of the survey.  We can consider $\phi$ as the (inverse) normalization constant since $\int \int P(L,z) dL dz \equiv 1$, defining $1/\phi_0 = N$.


For the entire sample of $N$ objects, the probability of $L$ and $z$ is simply the product of the individual probabilities:
\begin{equation}
P(L,z|\alpha,\beta,\phi_0) \propto \prod_{i=i}^{N} \frac{1}{N} \frac{dV}{dz}(z_i) \left(\frac{L_i}{L_0}\right)^{-\alpha}\left(\frac{1+z_i}{1+z_0}\right)^{\beta}.
\label{eqn:2}
\end{equation}


However, we do not observe the full sample due to incompleteness in e.g. flux.  We do understand our selection function $S$ and have a subsample of $n$ objects; the observations are a binomial process of $n$ draws from an intrinsic distribution containing $N$ objects.  Hence we can write

\begin{equation}
P(N|n,\alpha,\beta) = P(N) C^{N}_{n}\left(P(objects~in~sample|\alpha,\beta)\right)^n\left(P(objects~not~in~sample|\alpha,\beta)\right)^{N-n},
\end{equation}
which is equal to
\begin{equation}
P(N|n,\alpha,\beta) = C^{N-1}_{n-1}\left(\int\int S(L,z)P(L,z|\alpha,\beta) dLdz\right)^n
\left(1-\int\int S(L,z)P(L,z|\alpha,\beta) dLdz\right)^{N-n},
\label{eqn:5}
\end{equation}
where $n$ is the number of observations in our sample and we assume a flat prior on $log~N$.

The posterior on the parameters $\alpha$ and $\beta$ given the observed data set is simply the product of the $n$ observed probabilities:
\begin{equation}
P(\alpha,\beta|\{L_{obs},z_{obs}\}) = P(\alpha,\beta) \prod_{i=1}^n P_{obs}(L_i,z_i|\alpha,\beta)
\end{equation}
or
\begin{equation}
P(\alpha,\beta|\{L_{obs},z_{obs}\}) =\prod_{i=1}^n \frac{P(L_i,z_i|\alpha,\beta)}{\int\int S(L,z)P(L,z|\alpha,\beta) dLdz},
\end{equation}
which simplifies to
\begin{equation}
P(\alpha,\beta|\{L_{obs},z_{obs}\}) =\left(\int\int S(L,z)P(L,z|\alpha,\beta) dLdz\right)^{-n} \prod_{i=1}^n P(L_i,z_i|\alpha,\beta).
\label{eqn:3}
\end{equation}

We would like to know the posterior probability of the model parameters given our $n$ observed data points.  From Bayes's theorem, 

\begin{equation}
P(\alpha,\beta,N|\{L_{obs},z_{obs}\})\propto P(N|n,\alpha,\beta)\times P(\alpha,\beta|\{L_{obs},z_{obs}\}).
\end{equation}

Since the first term in Equation \ref{eqn:3} will cancel with the second term in Equation \ref{eqn:5}, we obtain a simplified posterior probability of


\begin{equation}
P(\alpha,\beta,N|\{L_{obs},z_{obs}\})\propto C^{N-1}_{n-1}\left(1-\int\int \frac{1}{N}\frac{dV}{dz}(z) \Phi(L,z|\alpha,\beta) S(L,z)dLdz\right)^{N-n} \prod_{i=1}^n P(L_i,z_i|\alpha,\beta).
\end{equation}

From here, a standard MCMC analysis can determine the distributions for the model parameters.

\section{Appendix C: Table of High-EW [O III] and H$\alpha$ Emitters}
\label{sec:table}
\LongTables
\begin{deluxetable*}{lccccccc}
\tablecaption{High-EW [O III] and H$\alpha$ Emitters}
\tablehead{ \colhead{ID} & \colhead{RA} & \colhead{Dec} & \colhead{$z_{grism}$} & \colhead{H$\alpha$ Flux} & \colhead{H$\alpha$ EW} & \colhead{[O III] Flux} & \colhead{[O III] EW}}
\startdata
AEGIS-1312 &        215.10797 &        52.931831 & 1.90 & ... & ... & 6.12 $\pm$ 0.386 & 563. $\pm$ 68.5\\
AEGIS-2157 &        215.00920 &        52.866520 & 1.17 & 3.83 $\pm$ 0.408 & 754. $\pm$ 149. & 1.21 $\pm$ 0.412 & 201. $\pm$ 83.1\\
\enddata
\label{tab:grism}
\tablecomments{ID numbers come from \citet{skelton}; all EW values are quoted in the restframe (\AA); fluxes are in units of 10$^{-17}$ erg s$^{-1}$ cm$^{-2}$. (The full table will be included in the published version)}
\end{deluxetable*}

\begin{deluxetable*}{lccccc}
\tablecaption{Plausible High-EW [O III] Emitters}
\tablehead{ \colhead{ID} & \colhead{RA} & \colhead{Dec} & \colhead{Observed Wavelength (\AA)} & \colhead{Line Flux} & \colhead{Line EW}}
\startdata
AEGIS-1342 &        215.09169 &        52.920654 & 12170 & 1.52 $\pm$ 0.393 & 2580 $\pm$ 2220\\
AEGIS-2591 &        214.82121 &        52.734825 & 14870 & 2.43 $\pm$ 0.565 & 1290 $\pm$ 537.\\
\enddata
\label{tab:plausiblegrism}
\tablecomments{ID numbers come from \citet{skelton}; all EW values are quoted in the observed frame (\AA); fluxes are in units of 10$^{-17}$ erg s$^{-1}$ cm$^{-2}$. (The full table will be included in the published version)}
\end{deluxetable*}
\end{document}